\pdfoutput=1
\documentclass[fullpage]{article}
\usepackage{graphicx}
\usepackage{amsmath}
\usepackage{afterpage}

\usepackage{tikz} 
\usepackage{caption}
\usepackage{subfigure}
\usepackage{multirow}
\usepackage{xcolor}

\usepackage{listings}

\usepackage{placeins}

\usetikzlibrary{arrows}
\usetikzlibrary{shapes}

\tikzset{
    *|/.style={
        to path={
            (perpendicular cs: horizontal line through={(\tikztostart)},
                                 vertical line through={(\tikztotarget)})
            -- (\tikztotarget) \tikztonodes
        }
    }
}

\tikzset{every picture/.style={line width=1pt}}

\definecolor{columbiablue}{rgb}{0.61, 0.87, 1.0}
\definecolor{classicrose}{rgb}{0.98, 0.8, 0.91}
\definecolor{lightgray}{rgb}{0.83, 0.83, 0.83} 
\definecolor{inchworm}{rgb}{0.7, 0.93, 0.36}
\definecolor{backcolor}{rgb}{0.95,0.95,0.92}

\lstdefinestyle{mystyle}{
    backgroundcolor=\color{backcolor},
    basicstyle=\ttfamily\footnotesize,
    breakatwhitespace=false,
    breaklines=true,
    captionpos=b,
    keepspaces=true,
    showspaces=false,
    showstringspaces=false,
    showtabs=false,
    tabsize=4
}

\begin{document}

\title{Parametric Differential Machine Learning for Pricing and Calibration}
\author{Arun Kumar Polala, Bernhard Hientzsch}
\maketitle

\begin{abstract}
Differential machine learning (DML) is a recently proposed technique that uses samplewise state derivatives to regularize least square fits 
to learn conditional expectations of functionals of stochastic processes as functions of state variables. 
Exploiting the derivative information leads to fewer samples than a vanilla ML approach for the same level of precision. 
This paper extends the methodology to parametric problems where the processes and functionals also depend on model and contract parameters, respectively.
In addition, we propose adaptive parameter sampling to improve relative accuracy when the functionals have different magnitudes for different parameter sets. 
For calibration, we construct pricing surrogates for calibration instruments and optimize over them globally. 
We discuss strategies for robust calibration. 
We demonstrate the usefulness of our methodology on one-factor Cheyette models with benchmark rate volatility specification with an 
extra stochastic volatility factor on (two-curve) caplet prices at different strikes and maturities, first for parametric pricing, 
and then by calibrating to a given caplet volatility surface.
To allow convenient and efficient simulation of processes and functionals and in particular the corresponding computation of 
samplewise derivatives, we propose to specify the processes and functionals in a low-code way close to mathematical notation 
which is then used to generate efficient computation of the functionals and derivatives in TensorFlow.  
\end{abstract}

\section{Introduction}
In many financial applications, the goal is to compute the conditional expectation $E[Y|X]$, 
given $N_S$ joint samples $\{ \left(X^i,Y^i\right), i = 1, ..., N_S \}$ 
with $X^i = \left( X^{i}_{j}, j = 1, ..., N_X \right)$ and $Y^i = \left( Y^{i}_{j}, j = 1, ..., N_Y \right)$ where $N_X$ and 
$N_Y$ are the dimensions of $X$ and $Y$, respectively. 
Often, $X$ is a random input to some stochastic computation (or a random intermediate result of such computation) and 
$Y$ is the output of that computation. 
In particular, $X$ could be the initial (or intermediate) value of a stochastic process $S$ while 
$Y$ is a function of the final state of that process (or a functional of the full trajectory of that process). 

Recently, machine learning techniques have been used to approximate $e(X)=E[Y|X]$ given enough joint samples of $X$ and $Y$. 
Let $N(X;\Theta_N)$ denote a deep neural network (DNN) trained to approximate $e(X)$. 
Here, $\Theta_N$ denotes the trainable DNN parameters (typically weights and biases). 
The standard approach to train the best DNN is to minimize least squares as follows:         
\begin{equation}\label{VMLlosseqn}
\Theta_N^{*,ML} =  \underset{\Theta_N}{\arg\min} E\left[\left| Y - N(X;\Theta_N) \right|^2 \right],  
\end{equation}
resulting in $e^{*,ML}(X):= N\left(X;\Theta_N^{*,ML}\right)$. 
The standard approach is usually referred as vanilla ML approach in the literature. 
Recently, Huge and Savine \cite{huge2020differential,huge2020differentialrisk} proposed a technique called Differential Machine Learning (DML) that uses  
information from sample-wise differentials to improve on the vanilla ML approach. 
Specifically, it minimizes least squares subject to regularization by derivatives as follows:
\begin{equation}\label{DMLlosseqn}
\Theta_N^{*,DML} =  \underset{\Theta_N}{\arg\min} E\left[\left| Y - N(X;\Theta_N) \right|^2 
+ \sum_{i,j} \lambda_{ij} \left| DY_{ij} - \frac{\partial N(X;\Theta_N)_{i}} {\partial X_{j}} \right|^2 \right], 
\end{equation}
resulting in  $e^{*,DML}(X):= N\left(X;\Theta_N^{*,DML}\right)$, where $DY_{ij}=\frac{\partial Y_{i}}{\partial X_{j}}$. 
In the case where conditional expectation and partial derivatives commute, the DML approach has the same optimal parameter sets as vanilla ML approach but can be learned more sample efficiently\footnote{Sample efficiency here refers to obtaining same required accuracy with substantially fewer training samples.}. In \cite{huge2020differential,huge2020differentialrisk}, Huge and Savine considered initial or intermediate states as input vector $X$ and final payoffs as output vector $Y$, and showed that for various financial instruments, the DML approach was more sample efficient than vanilla ML approach, especially with respect to approximation of derivatives. 

We extend the DML technique to settings with parameters and call the technique Parametric Differential Machine Learning (PDML). 
The ``parametric" in PDML comes from including parameters of the stochastic computation in $X$. When we consider functionals of stochastic 
processes, these could be model parameters for the stochastic process (for example, Heston parameters) or parameters related to the functional
(for example, strike of (call) option) which we also call ``contract parameters".
Using PDML technique we can learn efficient DNN surrogates to approximate conditional expectation (aka price of a financial instrument) as a function of model and/or contract parameters which can be used to either price several financial instruments simultaneously or to calibrate model parameters against market data of financial instruments. 
Here, note that efficient application of PDML requires efficient computation of partial derivatives of DNN surrogate $N(X;.)$ and output vector $Y$ wrt input vector $X$. 
In Section \ref{DNNfmk_compgraph}, we review how we obtain partial derivatives of DNN surrogate $N(X;.)$ wrt input vector $X$ via twin networks. 
This leaves the efficient computation of $\frac{\partial Y}{\partial X}$. Existing approaches typically involve the use of algorithmic differentiation or
adjoint algorithmic differentiation (AD/AAD) techniques. 
Such AD/AAD approaches typically involve specifically prepared AD/AAD enabled source code which is then compiled and run in various modes, with the AAD system
extracting and recording computational graphs, which are then used to compute derivatives. 
However, working with such approaches is often cumbersome and involved. 
A limited application domain such as simulation of a class of general stochastic processes and computation of functionals allows one to describe the computations
to be performed (and differentiated) directly without executing them at first, to automatically generate the computations to be performed for any needed derivatives,
and to run these computations for values and derivatives efficiently.

In Section \ref{MLcondExp}, we describe a general set-up for these stochastic processes and functionals, how these stochastic prccesses and functionals can be specified by 
text input close to mathematical notation, and how the Python parser and TensorFlow 1 can be used to generate computational graphs for the computation of $Y$ and 
of   $\frac{\partial Y}{\partial X}$.  These computational graphs can be executed efficiently on CPU and GPU. 
In this way, we can efficiently and conveniently generate $Y$ and  $\frac{\partial Y}{\partial X}$ for a large set of stochastic processes and functionals. 

In Section \ref{applicationPDML}, we describe the set-up for the PDML, the learning of surrogates, and parametric pricing. We will discuss in particular how to achieve
good enough relative accuracy if magnitudes of prices for different parameter sets differ. 

In Section \ref{pricingsection}, we introduce the model - Cheyette Model with
benchmark rate volatility specification with an additional stochastic volatility factor, 
the instruments - two curve caplets, the parametric
pricing setup, and parametric pricing results both for uniform and adaptive parameter sampling approaches. 
In Section \ref{calibrationPDML}, we discuss how to calibrate by optimizing on DNN surrogates for single maturity and multi-maturity settings with global optimization methods. We also 
discuss how random numbers in various parts of the pipeline impact the quality of the surrogates and the optimized parameters obtained from those surrogates. Finally, we discuss how to use that
randomness, multiple seeds, and comparison to some ground truth indicators to obtain more robust calibration methods. In Section
\ref{calibrationresults}, we apply these calibration approaches and report
promising results for both simple and robustified calibration methods 
for single maturity and multi-maturity settings. In Section \ref{conclusion}, we conclude.  

The parametric pricer developed by the PDML technique can compute prices and greeks for different financial instruments with different parmeter sets simultaneously,
efficiently, and quickly. These parametric pricers allow much faster risk analytics computations of a portfolio of financial instruments 
and can be used as fast solvers inside a calibration approach. 

Often, parametric pricing starts with specialized fast implementations of often simplified models reduced 
to a small number of factors since that parametric pricer will be called very often in the pricing 
and calibration setting, in particular repeatedly for each evaluation of the calibration objective 
function. Those specialized solvers and formulations for those models as well 
as appropriate simplifications need to be derived and implemented and are problem-specific. In the PDML
demonstrated here, one only needs a differentiable implementation of the MC simulation and pricing 
under the original model. Instead of having to derive and implement each solver so that it fits 
within a particular API, we can construct such solvers based on input that looks very close to mathematical 
notation. The PDML technique results in DNN surrogates that can be very efficiently evaluated within 
the optimization and calibration loop so that now combining several surrogates and runs become feasible and 
the calibration can be further robustified. The proposed PDML technique gives promising results even 
as some of the components have not been separately optimized or adapted and thus further improvements 
are likely possible from future work. 

\section{Using DNNs to Represent Functions and Compute Derivatives}\label{DNNfmk_compgraph}

Deep neural networks (DNN) with nonlinear activation functions are universal function approximators, for a variety of architectures and some condition on the activation function. 
Modern deep learning frameworks, as for example TensorFlow and PyTorch, allow one to easily create and train DNNs of various architectures. One generally proceeds by specifying 
how one or several objective (``loss") function(s) can be computed by
(nontrainable) computations involving trainable DNNs and variables. In these frameworks, these computations are captured, either explicitly (as in TensorFlow 1 or TensorFlow 2's TensorFlow 1 compatibility mode) or implicitly (by decorating and/or analyzing Python operations and functions), as computational graphs. The operations allowed 
in computations in these frameworks have efficiently implemented derivatives (both forward and adjoint) with respect to their arguments. Using these derivatives of the operations, 
these frameworks implement well known algorithmic differentiation techniques \cite{naumann2011art} so that given the 
original computation, computations of appropriate derivatives can be easily added to the explicit or implicit computational graph.

Given these convenient and readily available frameworks, practioners have found that DNNs as parts of computational graphs are not only useful as theoretical constructs 
but show impressive results and efficient representations for a wide range of applications. The frameworks have implemented highly optimized computational infrastructure 
that can train on and evaluate such computational graphs and neural networks very efficiently on a wide range of computers and processors, from laptops over browsers running
Google Colab notebooks on shared Google Colab resources to on-premises high-end multi-GPU and multi-core multi-CPU servers.

As an example, a fully connected feedforward DNN can be described by the equations \cite[Section 1.1.1]{huge2020differential}, \cite{huge2020differentialrisk}:
\begin{eqnarray}
\nonumber
z_0 & = & x \\
z_{l} & = & W_{l} \rho_{l-1}\left(z_{l-1}\right) + b_{l} =: \mathbf{FwdLayer}_l(z_{l-1},W_l,b_l)\quad , l=1,\ldots,L \\
\nonumber
y & = & z_{L}
\end{eqnarray}
where the $\rho$ are the activation functions, the $W$ are the (matrices with) weights, and the $b$ are the (vectors of) biases. 
The activation function of the first layer or the last layer or for the final result $y=\rho_L(z_L)$
can be chosen differently\footnote{In our 
equations and computational graphs, we use identity $\rho_L(x)=x$. 
The equations and graphs can be easily adapted to the case where $\rho_L$ is not the identity.} than the ones for the middle layers. 
Typically, the same activation function is used for all middle layers. 
If there are constraints on the final output (such as that $y$ is supposed to be within a particular domain, such as positive,  
with a value between zero and one as for probabilities, or with components nonnegative and summing to one), these constraints
can be enforced by a particular choice of activation function to be applied after the last layer, $y=\rho_L(z_L)$, 
or in the last layer, $\rho_{L-1}$.  If there are no such constraints, identity is a 
common choice for the last two activation functions.

We note that it is best practice to scale the inputs $x$ to DNNs and otherwise preprocess the input \cite{huge2020differential}. 
Once determined, scaling can be implemented either before input is provided to the network, or as a particular 
first layer of the network with no or only some trainable elements. In particular cases, this can be implemented 
as a first layer with an identity activation function $\rho_0(x)=x$.  

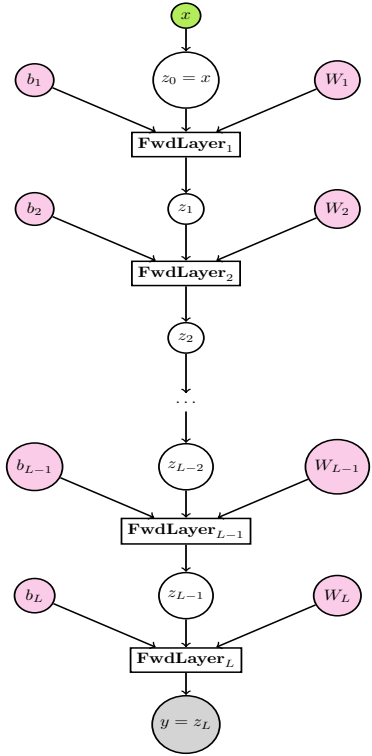
\begin{figure}[h!]
\begin{center}
\resizebox{50mm}{100mm}{
\begin{tikzpicture}

\node[draw,circle,fill=inchworm] (x) at (3,16.5) {$x$};
\node[draw,circle,fill=classicrose] (b1) at (0,15) {$b_1$};
\node[draw,circle] (z0) at (3,15) {$z_0=x$}; 
\node[draw,circle,fill=classicrose] (W1) at (6,15) {$W_1$};
\node[draw] (fwdlayer1) at (3,13.5) {$\mathbf{FwdLayer}_1$};

\draw[->] (x) to (z0);
\draw[->] (b1) to (fwdlayer1); 
\draw[->] (z0) to (fwdlayer1);
\draw[->] (W1) to (fwdlayer1);

\node[draw,circle,fill=classicrose] (b2) at (0,12) {$b_2$};
\node[draw,circle] (z1) at (3,12) {$z_1$}; 
\node[draw,circle,fill=classicrose] (W2) at (6,12) {$W_2$};
\node[draw] (fwdlayer2) at (3,10.5) {$\mathbf{FwdLayer}_2$};

\draw[->] (fwdlayer1) to (z1);
\draw[->] (b2) to (fwdlayer2); 
\draw[->] (z1) to (fwdlayer2);
\draw[->] (W2) to (fwdlayer2);

\node[draw,circle] (z2) at (3,9) {$z_2$}; 

\draw[->] (fwdlayer2) to (z2);

\node (noz) at (3,7.5) {$\mathbf{\cdots}$};

\draw[->] (z2) to (noz);

\node[draw,circle,fill=classicrose] (bl1) at (0,6) {$b_{L-1}$};
\node[draw,circle] (zl2) at (3,6) {$z_{L-2}$}; 
\node[draw,circle,fill=classicrose] (Wl1) at (6,6) {$W_{L-1}$};
\node[draw] (fwdlayerl1) at (3,4.5) {$\mathbf{FwdLayer}_{L-1}$};

\draw[->] (noz) to (zl2);
\draw[->] (bl1) to (fwdlayerl1); 
\draw[->] (zl2) to (fwdlayerl1);
\draw[->] (Wl1) to (fwdlayerl1);

\node[draw,circle,fill=classicrose] (bl) at (0,3) {$b_{L}$};
\node[draw,circle] (zl1) at (3,3) {$z_{L-1}$}; 
\node[draw,circle,fill=classicrose] (Wl) at (6,3) {$W_{L}$};
\node[draw] (fwdlayerl) at (3,1.5) {$\mathbf{FwdLayer}_{L}$};

\draw[->] (fwdlayerl1) to (zl1);
\draw[->] (bl) to (fwdlayerl); 
\draw[->] (zl1) to (fwdlayerl);
\draw[->] (Wl) to (fwdlayerl);

\node[draw,circle,fill=lightgray] (y) at (3,0) {$y=z_{L}$}; 

\draw[->] (fwdlayerl) to (y);

\end{tikzpicture}
}
\end{center}
\caption{Computational Graph for Feedforward Network \label{fwdnetwork}}
\end{figure}

To demonstrate, Figure \ref{fwdnetwork} shows the corresponding computational graph.  
Green circle nodes are inputs, gray circle nodes are outputs, pink circle nodes are trainable parameters (matrices and vectors), unfilled boxes are computations, 
and unfilled circle nodes are intermediate results.

Neural networks are typically trained through back-propagation, a version of adjoint algorithmic differentiation adapted to the form of the networks, by proceeding 
backwards through the layers, as in: 
\begin{eqnarray}
\nonumber
\bar z_{L} & = & \bar y = 1\\
\bar z_{l-1} & = & \left(\bar z_{l}W_l^T\right)\circ \rho_{l-1}^{\prime}\left(z_{l-1}\right)=:\mathbf{BwdLayer}_l({\bar z_l},z_{l-1},W_l) \quad , l=L,\ldots,1 \\
\nonumber
\bar x & = & \bar z_0
\end{eqnarray}
where the adjoint notation is used: $\bar x = \partial y / \partial x,\bar z_l = \partial y / \partial z_l, \bar y = \partial y / \partial y=1$, 
$\circ$ is the elementwise product, and $\rho_l^{\prime}$ is the derivative of the activation function \cite[Section 1.1.2]{huge2020differential}. 
For this derivative to be smooth and make computational sense, the activation functions $\rho_l$ have to be differentiable\footnote{Alternatively, one could look for 
functions $\rho_l^{\prime}$ that approximate the derivative well in a certain sense but this will lead to notational and other difficulties. Since networks with 
differentiable activation functions work well enough, we will not discuss such approaches here.}, excluding 
the often used ReLU. However, there is a rather large variety of such activation functions including sigmoid, ELU, softplus, or swish. 

\begin{figure}[h!]
\begin{center}
\resizebox{100mm}{100mm}{
\begin{tikzpicture}

\node[draw,circle,fill=inchworm] (x) at (3,16.5) {$x$};
\node[draw,circle,fill=classicrose] (b1) at (0,15) {$b_1$};
\node[draw,circle] (z0) at (3,15) {$z_0=x$}; 
\node[draw,circle,fill=classicrose] (W1) at (6,15) {$W_1$};
\node[draw] (fwdlayer1) at (3,13.5) {$\mathbf{FwdLayer}_1$};

\node[draw] (bwdlayer1) at (9,16.5) {$\mathbf{BwdLayer}_1$};
\node[draw,circle] (zb1) at (9,15) {$\bar z_1$};
\node[draw,circle] (zb0) at (12,16.5) {$\bar z_0$};

\draw[->] (x) to (z0);
\draw[->] (b1) to (fwdlayer1); 
\draw[->] (z0) to (fwdlayer1);
\draw[->] (W1) to (fwdlayer1);

\draw[->] (bwdlayer1) to (zb0);
\draw[->] (z0) to (bwdlayer1);
\draw[->] (W1) to (bwdlayer1);
\draw[->] (zb1) to (bwdlayer1); 

\node[draw,circle,fill=classicrose] (b2) at (0,12) {$b_2$};
\node[draw,circle] (z1) at (3,12) {$z_1$}; 
\node[draw,circle,fill=classicrose] (W2) at (6,12) {$W_2$};
\node[draw] (fwdlayer2) at (3,10.5) {$\mathbf{FwdLayer}_2$};

\node (nozb) at (9,10.5) {$\mathbf{\cdots}$};

\node[draw] (bwdlayer2) at (9,13.5) {$\mathbf{BwdLayer}_2$};
\node[draw,circle] (zb2) at (9,12) {$\bar z_2$};

\draw[->] (fwdlayer1) to (z1);
\draw[->] (b2) to (fwdlayer2); 
\draw[->] (z1) to (fwdlayer2);
\draw[->] (W2) to (fwdlayer2);

\draw[->] (bwdlayer2) to (zb1);
\draw[->] (W2) to (bwdlayer2);
\draw[->] (zb2) to (bwdlayer2);
\draw[->] (z1) to (bwdlayer2); 
\draw[->] (nozb) to (zb2);

\node[draw,circle] (z2) at (3,9) {$z_2$}; 

\node[draw,circle] (zbl2) at (9,9) {$z_{L-2}$}; 

\draw[->] (fwdlayer2) to (z2);
\draw[->] (zbl2) to (nozb);

\node (noz) at (3,7.5) {$\mathbf{\cdots}$};

\node[draw] (bwdlayerl1) at (9,7.5) {$\mathbf{BwdLayer}_{L-1}$};

\draw[->] (z2) to (noz);
\draw[->] (bwdlayerl1) to (zbl2);

\node[draw,circle,fill=classicrose] (bl1) at (0,6) {$b_{L-1}$};
\node[draw,circle] (zl2) at (3,6) {$z_{L-2}$}; 
\node[draw,circle,fill=classicrose] (Wl1) at (6,6) {$W_{L-1}$};
\node[draw] (fwdlayerl1) at (3,4.5) {$\mathbf{FwdLayer}_{L-1}$};

\node[draw,circle] (zbl1) at (9,6) {$\bar z_{L-1}$};

\draw[->] (noz) to (zl2);
\draw[->] (bl1) to (fwdlayerl1); 
\draw[->] (zl2) to (fwdlayerl1);
\draw[->] (Wl1) to (fwdlayerl1);

\draw[->] (zbl1) to (bwdlayerl1);
\draw[->] (zl2) to (bwdlayerl1);
\draw[->] (Wl1) to (bwdlayerl1);

\node[draw,circle,fill=classicrose] (bl) at (0,3) {$b_{L}$};
\node[draw,circle] (zl1) at (3,3) {$z_{L-1}$}; 
\node[draw,circle,fill=classicrose] (Wl) at (6,3) {$W_{L}$};
\node[draw] (fwdlayerl) at (3,1.5) {$\mathbf{FwdLayer}_{L}$};

\node[draw] (bwdlayerl) at (9,4.5) {$\mathbf{BwdLayer}_L$};

\node[draw,circle] (one) at (9,3) {$1$};

\draw[->] (fwdlayerl1) to (zl1);
\draw[->] (bl) to (fwdlayerl); 
\draw[->] (zl1) to (fwdlayerl);
\draw[->] (Wl) to (fwdlayerl);

\draw[->] (bwdlayerl) to (zbl1);
\draw[->] (zl1) to (bwdlayerl);
\draw[->] (Wl) to (bwdlayerl);
\draw[->] (one) to (bwdlayerl);

\node[draw,circle,fill=lightgray] (y) at (3,0) {$y=z_{L}$}; 

\node[draw,circle,fill=lightgray] (dy) at (9,0) {$\frac{\partial y}{\partial x} = \bar z_0$};

\draw[->] (fwdlayerl) to (y);
\draw[->] (zb0) to[bend left] (dy); 

\end{tikzpicture}
}
\end{center}
\caption{Computational Graph for Feedforward Network with Added Adjoint Computation of
Derivative (``Twin network") \label{twinnetwork} }
\end{figure}
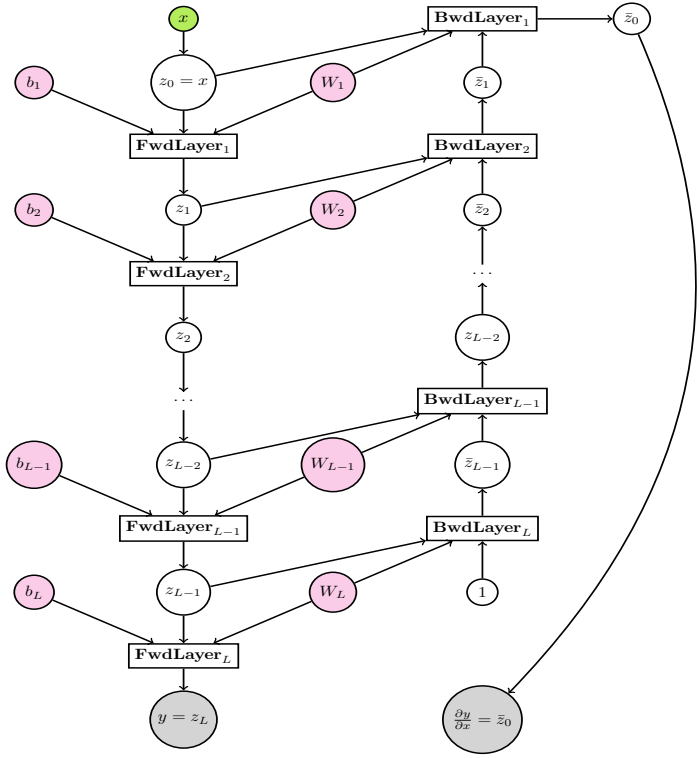

At the end of the backward pass, $\bar x$ will contain the derivative of $y$ with respect to the input $x$. 
Just like the original network could be created easily in the framework, 
the second network computing the derivative of the network with respect to its input(s), 
can be created similarly, and one obtains a network that computes both the output and its 
derivative with respect to the input, in which the weights $W$ are shared between forward and backward network. 
This combined network is called ``twin network" by Huge and Savine. We show in
Figure \ref{twinnetwork} what nodes and computations are added to the forward 
network to also compute $\frac {\partial y }{ \partial x}$ by adjoint algorithmic differentiation.

Instead of explicitly implementing the backward network, 
one can also use the built-in functionalities of the framework 
to make the framework generate the computational graph for the and/or compute the specified derivatives. In TensorFlow, this is called 
tensorflow.gradient and adds operations to the computational graph to also compute the specified derivative, with respect to any input,
parameter, or intermediate result specified.  

In this way, we can easily obtain efficient implementations of the DNNs  and of needed derivatives with respect to inputs and parameters. 

To implement the forward mode of AD, one can extend the vectors so that they record both value and derivative of value with respect to input(s) 
and extend the operators so that they also correctly propagate the derivatives. 
  
\section{Differentiable Computation of Functionals of Simulated Stochastic Processes}\label{MLcondExp}

\subsection{General Setup}
In quantitative finance applications, one very often has a description (model) 
that describes the dynamics of a number of risk factors or variables over time, 
say $S_t$ (vector, for a range of $t$), 
being started at $S_0$ (vector). This dynamics has parameters called model parameters ($\Theta_S$). 
Then, one or several payoffs - or other quantities of financial interest - are defined as functionals $F(S_{[0,T]})$ 
that in general can take the entire path between time 0 and time $T$ (the ``maturity" or ``computing time" of the payoff or quantity) 
as input and return one output. This functional can have parameters as well (``contract parameters", $\Theta_F$). 
For example, consider an Asian call option in Black-Scholes model set-up:
\begin{eqnarray}
dS_t &=& \mu S_t dt + \sigma S_t dW_t, \\
F(S_{[0,T]}) &=& \left(  \int_0^T S_t dt - K \right)^+.
\end{eqnarray}
Here, $\mu, \sigma$ are model parameters from $\Theta_S$ and $K$ is a contract parameter from $\Theta_F$.

In case of path-dependent functionals the entire path needs to be saved and processed which is considerably harder than 
working with functionals that depend on the state at one or several given "fixing" times. 
So, for path-dependent functionals, one often looks for an exact or approximate
markovianization by  introducing an additional vector $U_t$ of ``extra state" 
which turns the functional into a markovian functional\footnote{
Or approximate the functional with a markovian one.} $F^U$, i.e., $F^U(S_T,U_T) := F(S_{[0,T]})$.
For example, for an Asian call option, one possibility would be:
\begin{eqnarray}
U_t &=& \int_0^t S_s ds, \\
F(S_{[0,T]}) &=& \left( U_T - K \right)^+, \\
&=& F^U(S_T,U_T).
\end{eqnarray}
Finally, $F(S_{[0,T]})$ respectively  $F^U(S_T,U_T)$ is sampled and used for instance to estimate 
expectations or conditional expectations.

The further set-up now depends on the particular application. For instance, we are interested in parametric pricing 
of a fixed instrument assuming a given fixed starting state\footnote{
Or some starting state, say corresponding to market 
situation within the next minutes or hours, $S_0$ within some specified interval.
} $S_0$.  Under those circumstances, the input vector $X$ 
in the PDML setup from the introduction would be only\footnote{
Adding $S_0$ as an additional element of $X$ 
for the case where starting value is not fixed.} the model parameters $\Theta_S$. Alternatively, one could try to learn a pricer across contract 
parameters also (varying strikes $K$, for instance, and then these contract parameters would be added as elements to $X$) 
and then use this to price a set of contracts of interest, such as calibration contracts. This is the parametric pricing set up that 
we will discuss here (including appropriate sensitivities).

Alternatively, for other applications, one could be interested in future values or pricing or exposures as of a 
certain time $T_1$, where $X$ would contain the future state $S_{T_1}$ (or $\left( S_{T_1}, U_{T_1} \right)$ for the markovianized case). 
Here one could potential simulate the process up to time $T_1$ according to the same or some other dynamics or sample 
the $S_{T_1}$ (or $\left( S_{T_1}, U_{T_1} \right)$ appropriately) and start the simulation at time $T_1$. This is the setting that 
Savine and Huge concentrate on. We will discuss such applications in future papers, but concentrate on parametric pricing 
here.

\subsection{A General Class of Stochastic Processes and Functionals}
In this section, we discuss the setup of system dynamics and its simulation. We begin with an assumption that we have specifications for a to-be-simulated/analyzed 
system as follows:\footnote{This specification is general enough for the purposes of this article, but it has been and is being extended for other applications and models, as described in other papers.}
Each component of the system is simulated or computed on some given time discretization grid either as
\begin{itemize}
\item A (presumably time-discretized) SDE/ODE with drift term (which can depend on all components and time) and volatility term(s) (which can depend on all components and time) 
associated with certain potentially correlated Brownian increments, with specified initial values (which could be fixed or generated according to some random distribution) OR
\item A deterministic function of time and only already computed components at the same time  OR
\item A deterministic update function with specified initial values, whose next value is given as a function of its value at the previous time step, time corresponding to current time step,
size of the current time step, potentially values of all other components at the previous time step, and values of only already computed components at the current time step. 
\end{itemize}
The specification requires a set of potentially correlated Brownian increments. Correlations between any two (groups of) 
Brownian increments can be given as any functions of any components at the beginning of the corresponding time step.

With formulas, this looks as follows: Denoting component $i$ by $x^i$, allowing $K_i=0$ (in which case the sum is defined to be zero),
allowing the drift functions to be zero, denoting ${\cal I}_S$ the indices for components given by SDEs or ODEs, 
${\cal I}_F$ the indices for components given by functions, and $\mathsf{ind}(i,k)$ giving the index of the associated 
Brownian increment for the $k$th volterm for the $i$th component (volterms for different components can share the same Brownian increment), 
we have for the SDEs/ODEs, functions, and correlations of Brownians: 
\begin{eqnarray}
dx_t^i & = & \mathsf{drift}_i(t,(x_t^j)^M_{j=1}) dt + \sum_{k=1}^{K_i} \mathsf{vol}_{ik}(t,(x_t^j)^M_{j=1}) dW_t^{\mathsf{ind}(i,k)} \qquad (i \in {\cal I}_S)\\
x_t^i & = & f_i (t,(x_t^j)^{i-1}_{j=1}) \qquad (i \in {\cal I}_F)\\
\mathsf{corr}( dW_t^{i} , dW_t^{j} ) & = & \mathsf{correl}_{i,j} (t,(x^k_t)^M_{k=1}) 
\end{eqnarray}
For any $i,j$ without given $\mathsf{correl}_{i,j}$ function, we use $\mathsf{correl}_{i,j}={1}_{i==j}$ (diagonal entries are one, off-diagonal entries are zero) to 
complete the specification so that trivial entries do not need to be given. 

Update functions are defined on a discrete time grid with $t_{new}=t+\Delta t$ and  ${\cal I}_U$ are the indices for components given by update functions:
\begin{equation}
x^i_{t_{new}}  = \mathsf{upd}_i(t,\Delta t, (x^j_t)^M_{j=1},(x^j_{t_{new}})^{i-1}_{j=1})  \qquad (i \in {\cal I}_U)
\end{equation}

The sets of indices ${\cal I}_S$, ${\cal I}_F$, and ${\cal I}_U$ specify whether a given component is given as SDE (including ODE), markovian function, or 
update function. 

Using Euler-Maruyama time-stepping\footnote{Other general time-discretization methods can also be implemented similarly.}, 
$\Sigma$ and $\Delta W$ being 
appropriate matrices and vectors, respectively, $N(.,.)$ being the normal distribution with a given mean vector and correlation matrix, we obtain: 
\begin{eqnarray}
\Sigma^{ij} & = & \mathsf{correl}_{i,j} (t,(x^k_t)^M_{k=1}) \\
\Delta W_t & = &  \sqrt{\Delta t}  N(0,\Sigma) \\
x^i_{t_{new}} & = & x_t^i + \mathsf{drift}_i(t,(x_t^j)^M_{j=1}) \Delta t + \sum_{k=1}^{K_i} \mathsf{vol}_{ik}(t,(x_t^j)^M_{j=1}) \Delta W_t^{\mathsf{ind}(i,k)} \qquad (i \in {\cal I}_S) \\
x^i_{t_{new}} & = & f_i (t,(x^j_{t_{new}})^{i-1}_{j=1}) \qquad (i \in {\cal I}_F)\\
x^i_{t_{new}} & = & \mathsf{upd}_i(t,\Delta t, ((x_t^j)^M_{j=1},(x^j_{t_{new}})^{i-1}_{j=1})  \qquad (i \in {\cal I}_U) 
\end{eqnarray}
This can also be interpreted by replacing differentials by differences (for $dx^i_t$ and $dt$) 
and by replacing Brownian increments $dW_t^{\mathsf{ind}(i,k)}$ by appropriately correlated normal random numbers $\Delta W_t^{\mathsf{ind}(i,k)}$ in the SDE/ODE 
expressions.

Finally, we intend to compute or estimate conditional expectations, which could be fully path-dependent and include a payoff $P$, a numeraire  $\mathsf{Num}$, and/or 
a discount factor  $\mathsf{DF}$ 
\begin{equation}
   E\left[ F( x_{[0:T]}) | \cdot \right] = E\left[ \frac {P( x_{[0:T]})}{\mathsf{Num}(x_{[0:T]})} | \cdot \right] = 
   E\left[{P( x_{[0:T]})}{\mathsf{DF}(x_{[0:T]})} | \cdot \right], 
\end{equation}
or depend only on components at certain sampling times
\begin{equation}
  E\left[ \frac {P^S( (x_{T_k})^K_{k=1})}{\mathsf{Num}^S((x_{T_k})^K_{k=1})} | \cdot \right] =
  E\left[ {P^S( (x_{T_k})^K_{k=1})}{\mathsf{DF}^S((x_{T_k})^K_{k=1})} | \cdot \right], 
\end{equation}
or could be completely markovian
\begin{equation}
  E\left[ \frac {P^U(x_T)}{\mathsf{Num}^U(x_T)} | \cdot \right] =
  E\left[ {P^U(x_T)}{\mathsf{DF}^U(x_T)} | \cdot \right]. 
\end{equation}  

Our script framework discussed in Section \ref{scriptsection} allows simple specification of the sampling times settings and markovian setting for the conditional expectations, with a syntax inspired by \cite{savine2021modern}. 

\subsection{Conventional Approaches for Simulating Stochastic Processes and Computing Functionals and Their Derivatives}

In most quantitative finance implementations of Monte-Carlo pricers, often models are implemented ad-hoc and one-off, 
with specialized implementations for each model, typically in C++, and then combined with appropriate payoffs 
for pricers, see  \cite{joshi2004c++} for an introduction. To obtain differentiable implementations, one needs
to write appropriately structured and prepared C++ code so that it can be combined with some Algorithmic (Adjoint) 
Differentiation (AD/AAD) framework or one's own implementation of AAD, as discussed very well in \cite{savine2018modern}, including
the modern use of Expression Templates. 

Industry leading implementations have started to allow the user to describe 
payoffs in a script-like form, see \cite{savine2021modern}, and process these scripts into object representations that can be 
used for simulation and pricing but also for more advanced applications such as computing CVA from the same 
script input.

However, we are not aware of any implementations in which the models (i.e., the stochastic processes) can be scripted and 
still be simulated efficiently.

Also, the preparation and maintenance of such appropriately structured C++ places a substantial burden on quantitative developers. 
Developers also need to use this instrumented code appropriately to extract, record, and process computational graphs (``tapes") 
through their AAD frameworks. 
The framework is then used to compute needed derivatives based on those processed graphs, which requires additional care and burden. 
In the next subsection, we will discuss a much more convenient and automatic way to do so for the class of processes and 
functonals that we just described.

\subsection{Script Framework for Simulating Stochastic Processes and Functionals}\label{scriptsection}

To the best of our knowledge, we are the first to implement model scripting. 
We started with a C++ framework and examined its usefulness for standard and hybrid models 
(see \cite{ogetbil2020extensions,ogetbil2020calibrating,ogetbil2022calibrating,ogetbil2022flexible}). 
This C++ framework allows the user to freely combine separately 
implemented sub-SDE systems, markovian updates, some functions and formulas, and payoffs. 
It was driven from Python input and enhanced by the Python interface. While very 
powerful, implementing completely new models required implementations of new classes. The use of AD/AAD
frameworks allowed the computation of derivatives (albeit sometimes memory bound), but added complexity
and maintenance burden, as discussed above.

For our current work, we are using a framework in which the system to be simulated is described in close-to-mathematical formulation, the 
Python parser is selectively used to generate appropriate abstract syntax trees and generate efficient implementations 
for TensorFlow and Numpy. Any necessary derivatives can be added in  the TensorFlow implementation by standard TensorFlow capabilities.

As a first example, the input in Figure \ref{script1} specifies a log-Euler full truncation scheme for the Heston model to price a call option.\footnote{Variables 
not given or defined in the script 
have to be defined within a provided dictionary or in the surrounding Python code (shortrate, kappa, 
longtermvariance, volofvol, rho, initiallogspot,initialvariance, and maturity, for this particular script input).} 
  
\begin{figure}[h]  
\begin{lstlisting}
#system 
d_logstock = (shortrate-0.5*variance)*d_t+volatility*d_Z
d_variance = kappa * (longtermvariance-positivepart(variance))*d_t \
             +volofvol*volatility*d_W 
volatility = sqrt(positivepart(variance))

#correlations 
d_W*d_Z=rho

#initial values
init: logstock = initiallogspot
init: variance = initialvariance

#payoff
maturity: calloption pays (positivepart(exp(logstock[t])- strike)) \
          discountby exp(-shortrate*t)
\end{lstlisting}
\caption{Script Input for Log-Euler Full Truncation Scheme for Call Option under Heston Model \label{script1}} 
\end{figure}

The next listing in Figure \ref{script2} shows pricing of three call options: one on a continuous Asian average, one 
on a discrete Asian average, and an up-and-out barrier one with observation at each time step.
The model is still Heston, but now treated with an Euler full truncation scheme. 
This listing also demonstrates some features including markovian updates for time-integral and maximum across time.

\begin{figure}[h] 
\begin{lstlisting}
#system 
d_stock = stock*(shortrate*d_t+volatility*d_Z)
d_stockint = stock * d_t 
stockmax = max(stockmax,stock_new)
d_variance = kappa * (longtermvariance-positivepart(variance))*d_t \
             +volofvol*volatility*d_W 
volatility = sqrt(positivepart(variance))

#correlations 
d_W*d_Z=rho

#initial values
init: stock = initialspot
init: stockmax = initialspot
init: variance = initialvariance

#payoff
maturity: asiancalloption pays (positivepart(stockint[t]- asianstrike)) \
          discountby exp(-shortrate*t)         
maturity: asiancalloption1 pays (positivepart(0.25*(stock[0.25*t]+\
                                                    stock[0.5*t]+\
                                                    stock[0.75*t]+\
                                                    stock[t])-asianstrike) \
          discountby exp(-shortrate*t)          
maturity: uocalloption pays (positivepart(stock[t]-strike) \ 
                              if (stockmax[t]<barrier)     \
                              else zeroslike(stock[t]))    \
          discountby exp(-shortrate*t)
\end{lstlisting}
\caption{Script for Euler Full Truncation Scheme under Heston for Call Option on Continuous Asian Average, 
Call Option on Discrete Asian Average, and Up-and-Out Call Option with Barrier Observation at Each 
Time Step, Demonstrating Markovian Updates for Time-integral and Maximum.\label{script2}} 
\end{figure}

As mentioned above, our Tensorflow backend is able to generate fully differentiable TensorFlow computational graphs 
from this input for the simulation of the system and the payoffs. For debugging, tests, and demonstrations, 
we also provide a Numpy backend for direct simulation of the system and payoffs under Numpy. 

This framework has enabled us to very quickly and efficiently work with a wide range of models and payoffs 
based on relatively minimal input in a form close to mathematical notation. 
  
\section{Using PDML for Computing Conditional Expectations}\label{applicationPDML}

\subsection{Training of PDML Network Using Differentials wrt Parameters}

As discussed in the introduction, there are various approaches to train DNNs for conditional expectations based
on samples.

Without differential regularization, we obtain the ``Vanilla ML" regression with loss function (see equation (\ref{VMLlosseqn})): 

\begin{equation*}
\Theta_N^{*,ML} =  \underset{\Theta_N}{\arg\min} E\left[\left| Y - N(X;\Theta_N) \right|^2 \right].  
\end{equation*}

With complete differential regularization as in the introduction, we obtain the (P)DML regression with loss fuction (see equation (\ref{DMLlosseqn})):
\begin{equation*}
\Theta_N^{*,DML} =  \underset{\Theta_N}{\arg\min} E\left[\left| Y - N(X;\Theta_N) \right|^2 
+ \sum_{i,j} \lambda_{ij} \left| DY_{ij} - \frac{\partial N(X;\Theta_N)_{i}} {\partial X_{j}} \right|^2 \right]. 
\end{equation*} 
This regularization with the sample-wise differentials significantly improves the training and typically much fewer samples are needed for the same accuracy,
even more so if a certain accuracy in derivative approximation is required. 
We use sample-wise differentials with respect to parameters (model, contract) and/or to state vector in training.

Sometimes, it is not necessary to include regularization with respect to all parameters or state features to achieve such improvement. 
Thus, we are using a setting where we allow partial differential regularization in the loss function: 
\begin{equation}\label{PDMLlosseqn}
\Theta_N^{*,PDML} =  \underset{\Theta_N}{\arg\min} E\left[\left| Y - N(X;\Theta_N) \right|^2 
+ \sum_{i,j} \lambda_{ij} \left| DY_{ij} - \frac{\partial N(X;\Theta_N)_{I(i)}} {\partial X_{J(j)}} \right|^2 \right]. 
\end{equation}
Here, $DY_{ij}=\frac{\partial Y_{I(i)}}{\partial X_{J(j)}}$. 
The mappings $I$ and $J$ specify which partial derivatives are used in the regularization. 

The output $Y$ and sample-wise differentials $DY_{ij}$ are obtained from the script framework
discussed in Section \ref{scriptsection}, which implements convenient derivatives with respect 
to parameters and initial or immediate state. 

We then learn DNN with appropriate deep learning approaches according to the corresponding loss functions by minimization.

\subsection{Learning the Surrogate}
As discussed above, $X$ could contain varying parameters, initial values, intermediate values, or other inputs
or outputs, coming from some simulated dynamics that could be started from fixed or varying initial values. 
$Y$ in general is a random variate -- typically some output of a given computation -- for which we want to compute a conditional expectation.
For parameters and initial values, we are typically given ranges or domains, and we would pick particular values for the
different realizations either by some deterministic scheme, quasi-random sampling, or random sampling according to some 
distributions. 

For intermediate values and outputs, appropriate ranges and domains can sometimes only be determined from simulating 
and sampling, and it might be difficult to make sure that the simulated intermediate values and outputs appropriately 
cover the domain of interest. If the system can be markovianized and all state in $X$ is intermediate state at $t$, 
one could instead start simulation from an appropriately described and constrained markovian state $M_t$ at 
the intermediate time $t$ in question.\footnote{If the system can be controlled with an additional drift, one could 
try to determine a drift adjustment so that a certain domain is reached or sampled, together with an appropriate 
measure change, within a similar framework.} Since in the setting in this paper, we do not condition on 
intermediate states, we will not discuss this further. 
 
In case components of $X$ are parameters, initial values, or intermediate values which in the computational graph are connected with 
the output $Y$ and come before $Y$, adjoint algorithmic differentiation and appropriately prepared forward
algorithmic differentiation can generate sample-wise derivatives of $Y$ with respect to those components of $X$. 
(If components of $X$ are outputs, one would need a reformulation or alternative approaches to compute partial derivatives 
of $Y$ with respect to those components of $X$. Since we are not handling this setting in this paper, we will 
not discuss this case further.)

We consider the case where components of $X$ are (dynamics or contract) parameters or potentially initial values. We 
generate computational graphs for the simulation and the computation of the functional(s). We will add nodes to the computational
graph to compute derivatives with respect to some or all components of $X$ with the TensorFlow backend. 

We will then generate vectors/tensors of appropriately deterministically or randomly sampled parameters and potentially initial
values and we will then use the computational graph to generate simulations of system trajectories from appropriately 
randomly generated Brownian increments for the SDEs and compute samples of $Y$ and $DY$ for these trajectories. We typically 
generate all needed realizations in one run, but one could generate more realizations as needed, potentially adaptively. 

We will specify the architecture of the surrogate DML network, such as a feedforward fully connected neural network with a 
certain number of hidden layers with a certain number of neurons each (but could use any other architecture that can be 
specified, initialized, and efficiently trained). Then, the weights, biases, or other trainable variables in the DML
network have to be appropriately (randomly) initialized. 

Then, we will use an appropriate optimization method to implement \eqref{DMLlosseqn} (or \eqref{VMLlosseqn} for VML).  This could be 
a stochastic gradient method with momentum or other features, such as mini-batch ADAM \cite{kingma2014adam},
could be a more conventional gradient-based
optimization with inexact or exact gradients such as L-BFGS, or could be optimization methods that treat the last layer explicitly 
by least-square methods as in \cite{cyr2020robust,cyr2019robustarxiv}. In our work reported here, mini-batch SGD
with ADAM with standard parameters seems to work well enough and so we leave more advanced methods to future work.  
Many of these optimization methods use random numbers.

Then, the optimization method is stoppped according to some stopping criterion, such as defined on changes to the parameters in the last 
optimization step, value of the objective function, or other considerations, and the resulting DML network is taken as surrogate.
Given that we are using random numbers in several parts of the process, the final surrogate might vary with those random numbers and
might be different and of different quality depending on the seed and other details of the random number generations. 

This entire process can be wrapped into an appropriate module and pipeline.    

\subsection{Addressing Different Magnitudes for $Y$ for Different Parameters - Adaptive Sampling}

If the $Y$ have all similar magnitude regardless of the parameter set, the above process will learn surrogates that fit 
the conditional expectation similarly well for parameters from different parts of the parameter domains. If the $Y$
vary in magnitude depending on the dynamics parameters, it turns out that DML surrogate will approximate the 
conditional expectations well for parameter settings for which the conditional expectations have the 
largest magnitude but not for parameters where the conditional expectations have smaller magnitudes.
This is often not desirable. 

There are many approaches to address this issue. For instance, one can sample the parameter regions resulting in smaller magnitude 
more intensively than others. To do this, one can first parametrically or nonparametrically estimate the magnitude of the $Y$ samples
over the parameters and then construct marginal (or bivariate) parameter distributions resulting in approximately constant expectation 
of $Y$ across the parameter domain. This approach seems to work well in our implementations and tests and its results are reported
below.

\section{Pricing Setting and Results}\label{pricingsection}

We examine pricing of caplets under Cheyette models with one interest rate factor $x$ (and thus one auxiliary factor $y$) and with 
an additional stochastic volatility factor $z$. Furthermore, we consider benchmark rate
volatility specifications for the ``local" volatility. Floorlets can be priced very similarly. 
Caps and floors can be priced as sums of caplets and floorlets, respectively. 
In our future work, we will consider European swaptions and also Cheyette models with several interest rate factors. 

\subsection{Pricing Model: Cheyette Model}

Here, we consider Cheyette models with a single interest rate factor with benchmark rate volatility specification 
\cite{andreasen2001turbo,andersenpiterbarginterest,schlenkrich2016quasigaussian}.
Our benchmark forward rate is the forward rate for tenor $\delta$. In this type of Cheyette model, the benchmark instantaneous forward rate is given as:
\begin{equation}
f(t,t+\delta) = f(0,t) + h(\delta)\left(x_t + y_t G(\delta) \right), \label{chfwdrate}
\end{equation}
where $h(\delta) = e^{-\kappa \delta}$ and  $G(\delta) = \frac{1 - h(\delta)}{\kappa}$ with $\kappa$ a given mean reversion. 

The interest rate state $x_t$ and the auxiliary factor $y_t$ follow the following SDEs and ODEs under the risk neutral (bank account) measure: 
\begin{eqnarray}
 dx_t &=& \left[y_t - \kappa x_t \right]dt + \sigma_{r}\left(t,x_t,y_t \right) dW^{Q}_t,  \\
dy_t &=& \left[\sigma_{r}^{2}\left(t, x_t, y_t \right) - 2 \kappa y_t \right] dt.   
\end{eqnarray}
Both $x$ and $y$ start at 0 at $t=0$, $x_0 \equiv 0$ and $y_0 \equiv 0$.

Under the $T$-Forward measure, the SDE for $x(t)$ acquires an additional drift adjustment: 
\begin{equation}
dx_t = \left[y_t - \kappa x_t - \sigma_{r}^{2}\left(t,x_t,y_t \right)G(T-t) \right]dt + \sigma_{r}\left(t,x_t,y_t \right) dW^{T}_t. 
\end{equation}

The ``local" volatility term is given by 
\begin{eqnarray}
\textbf{(no SV)}: \sigma_{r}\left(t,x_t,y_t\right) &=& h(\delta)\left[a(t)f(t,t+ \delta) + b(t) \right], \\
\textbf{(SV)}: \sigma_{r}\left(t,x_t,y_t\right) &=& \sqrt{z_t} h(\delta)\left[a(t)f(t,t+ \delta) + b(t) \right], 
\end{eqnarray}
which depends on $x_t$ and $y_t$ through (\ref{chfwdrate}).

In case of SV (stochastic volatility) setting, the dynamics of $z(t)$ is given by 
\begin{equation}
dz_t = \theta \left[z_{0} - z_t \right]dt + \eta(t) \sqrt{z_t}dZ_t, \hspace{2mm} z(0) \equiv z_{0} \equiv 1, \hspace{2mm} dZ_t dW^{Q (or T)}_t \equiv 0. 
\end{equation}

Under the Cheyette models under consideration, the discount factor curve as seen from time $t$ is given as:
\begin{equation}
P(t,T;x_t,y_t) = \frac{P(0,T)}{P(0,t)} \exp \left( -G(T-t) x_t -\frac{1}{2} G^2(T-t) y_t \right). 
\end{equation}

\subsection{Instruments: (Two-Curve) Caplets}

A caplet is an European style interest rate derivative. Let $T_{1} < T_{2}$, then the payoff of the caplet at $T_{2}$ for notional amount $N$ is
\begin{equation}\label{capletpayoff}
N \delta^C(T_1,T_2) \left(F_{T_{1}}^{F}\left(T_{1},T_{2}\right) - K\right)^+,
\end{equation}
where $K$ is the strike price, $F_{.}^{F}\left(T_{1},T_{2}\right)$ is the forward rate for the time period $\left[T_{1},T_{2}\right]$, $T_{1}$ is the reset date, 
$T_{2}$ is the payment date and $\delta^C$ is an appropriate day count fraction between $T_{1}$ and $T_{2}$ for the caplet payoff.

The forward rate can be computed from the discount factor for the forward/forecasting curve $P^F(t,T)$ - which represents the discount factor for $T$ as seen from $t$.
\begin{equation}
F_{t}^{F}\left(t^{S},t^{E}\right) = \frac{1}{\delta^F\left(t^{S},t^{E}\right)}\left(\frac{P^{F}\left(t, t^{S}\right)}{P^{F}\left(t, t^{E}\right)} - 1 \right),
\end{equation}
with $\delta^F$ being the appropriate day count fraction for the forecasting forward rate.

We have given stochastic models for the two curves (forecasting and discounting curves) 
in the form of discount factor functions $P^.(t,T;\mathsf{state}_t)$ where the $\mathsf{state}_t$ follows some 
stochastic process. In the Cheyette model under consideration,
$\mathsf{state}_t=(x_t,y_t)$ (no SV) or $\mathsf{state}_t=(x_t,y_t,z_t)$ (SV), but $P^.$ depends only on $(x_t,y_t)$ in both cases.
We are using models such as the Cheyette model that reproduce the original curves so that $P^.(0,T;\mathsf{state}_0)$ is equal to the curve $P^.(0,T)$ given as input. 

\subsubsection{Rewriting Two-Curve Payoffs into One-Curve Payoffs}

Below, we will rewrite two-curve payoffs, in particular caplet payoffs, into one-curve payoffs. In the two-curve setting, forecasting curve $P^{F}$ and the discounting curve $P^{D}$ will be different. To be able to rewrite two-curve payoffs into one-curve payoffs, we will need to make sufficient assumptions about the relationship
between the forecasting curve $P^{F}$ and the discounting curve $P^{D}$. One common such assumption is that these two curves have a deterministic multiplicative spread that 
does not change from the spread at time 0. This corresponds to hypothesis (\textbf{S0}) mentioned in \cite{MH_multicurve}.


Define
\begin{equation}
\beta_{t}^{F}(u,v)  := \frac{P^{F}(t,u)}{P^{F}(t,v)} \frac{P^{D}(t,v)}{P^{D}(t,u)}. 
\end{equation}
The deterministic multiplicative spread assumption corresponding to hypothesis \textbf{S0} assumes 
$\beta_{t}^{F}(u,v) = \beta_{0}^{F}(u,v)$.\footnote{Given our assumptions on the stochastic curve models,
 $\beta_{0}^{F}(u,v)$ can be computed from the given initial curves.}
This implies 
\begin{equation}
\frac{P^{F}(t,u)}{P^{F}(t,v)} = \beta_{0}^{F}(u,v) \frac{P^{D}(t,u)}{P^{D}(t,v)}.
\end{equation}
By setting $u=t$, we also have 
\begin{equation}
P^{D}(t,v) = \beta_{0}^{F}(t,v) P^{F}(t,v).
\end{equation}
We assume that $P^{D}$ and $P^{F}$ are using the same time convention and 
using hypothesis \textbf{S0} we rewrite the forward rate based on the forecasting curve as an affine function 
of the forward rate based on the discounting curve. 
\begin{eqnarray}
F_{t}^{F}\left(t^{S},t^{E}\right) &=& \frac{1}{\delta(t^S,t^E)}\left(\frac{P^{F}\left(t, t^{S}\right)}{P^{F}\left(t, t^{E}\right)} - 1 \right) \\
&=&  \frac{1}{\delta(t^S,t^E)}\left(\beta_{0}^{F}\left(t^{S},t^{E} \right) . \frac{P^{D}\left(t, t^{S}\right)}{P^{D}\left(t, t^{E}\right)} - 1 \right) \\
&=& \frac{1}{\delta(t^S,t^E)}\left(\beta_{0}^{F}\left(t^{S},t^{E} \right) \left( \frac{P^{D}\left(t, t^{S}\right)}{P^{D}\left(t, t^{E}\right)} - 1 \right) + \left(\beta_{0}^{F}\left(t^{S},t^{E} \right) - 1 \right) \right) \\
&=& \beta_{0}^{F}\left(t^{S},t^{E} \right)  \frac{1}{\delta(t^S,t^E)} . \left( \frac{P^{D}\left(t, t^{S}\right)}{P^{D}\left(t, t^{E}\right)} - 1 \right) + \frac{\beta_{0}^{F}\left(t^{S},t^{E} \right) - 1}{\delta(t^S,t^E)} \\
&=& \beta_{0}^{F}\left(t^{S},t^{E} \right) F_{t}^{D}\left(t^{S}, t^{E} \right) + \frac{\beta_{0}^{F}\left(t^{S},t^{E} \right) - 1}{\delta(t^S,t^E)} \\
&=& m F_{t}^{D}\left(t^{S}, t^{E} \right) + s,
\end{eqnarray}
with 
\begin{eqnarray}
 m & = &  \beta_{0}^{F}\left(t^{S},t^{E} \right) \\
 s & = & \frac{\beta_{0}^{F}\left(t^{S},t^{E} \right) - 1}{\delta(t^S,t^E)}. \\
\end{eqnarray}
If the two curves respective forward rates do not use the same time convention, $m$ will also contain the ratio of the day counts under the two different conventions.

\subsubsection{Two-Curve Caplet as One-Curve Payoff}
The caplet payoff given in \eqref{capletpayoff} at $T_2$ can be re-written, assuming that the day count for the caplet is the same as for forecasting forward rate and the same as for discounting forward rate (and thus denoted uniformly $\delta$):
\begin{eqnarray}
N \delta \left(F_{T_{1}}^{F}\left(T_{1},T_{2}\right) - K \right)^+  
&=& N \delta  \left(m F_{T_{1}}^{D}\left(T_{1},T_{2}\right) + s - K \right)^+\\
&=& N \delta  \left(\frac{m}{\delta} \left( \frac{1}{P^{D}(T_{1},T_{2})} - 1 \right) + s - K \right)^+  \\
&=& N \left(\frac{m}{P^{D}(T_{1},T_{2})} - m + (s-K)\delta \right)^+ \\
&=: & N \left(\frac{m}{P^{D}(T_{1},T_{2})} - \hat{K} \right)^+ , \label{twoasonepayoff}
\end{eqnarray}
with 
\begin{eqnarray*}
\hat{K} & = & m-s \delta + K \delta \\
& = & \beta_{0}^{F}\left(T_1,T_2 \right) - \frac{ \beta_{0}^{F}\left(T_1,T_2 \right) -1} {\delta} \delta + K \delta \\
& = & \beta_{0}^{F}\left(T_1,T_2 \right) - \left(\beta_{0}^{F}\left(T_1,T_2 \right) -1 \right) + K \delta \\
& = & 1 + K \delta .  
\end{eqnarray*}

We now simplify the first term in the parenthesis for the Cheyette Model:
\begin{eqnarray*}
 \frac{m}{P^{D}(T_{1},T_{2})} & = & \frac{m} {  \frac{P^D(0,T_2)}{P^D(0,T_1)} \exp \left( -G(T_2-T_1) x_{T_1} -\frac{1}{2} G^2(T_2-T_1) y_{T_1} \right)} \\
& = & \frac{m P^D(0,T_1)} {P^D(0,T_2)} \exp \left( G(T_2-T_1) x_{T_1} + \frac{1}{2} G^2(T_2-T_1) y_{T_1} \right). 
\end{eqnarray*}

The coefficient in front of the exponential function simplifies as follows:
\begin{eqnarray*}
\frac{m P^D(0,T_1)} {P^D(0,T_2)} & = & \frac{\beta_{0}^{F}\left(T_1,T_2 \right) P^D(0,T_1)} {P^D(0,T_2)} \\
& = & \frac{\left[\frac{P^{F}(0,T_1)}{P^{F}(0,T_2)} \frac{P^{D}(0,T_2)}{P^{D}(0,T_1)} \right] P^D(0,T_1)} {P^D(0,T_2)} \\
& = & \frac{P^{F}(0,T_1)}{P^{F}(0,T_2)}.
\end{eqnarray*}

Thus, (\ref{twoasonepayoff}) simplifies to:
\begin{eqnarray}
N \left(\frac{m}{P^{D}(T_{1},T_{2})} - \hat{K} \right)^+ & = &
N \left( \frac{P^{F}(0,T_1)}{P^{F}(0,T_2)}  \exp \left( G(T_2-T_1) x_{T_1} + \frac{1}{2} G^2(T_2-T_1) y_{T_1} \right)  - \hat{K} \right)^+  \nonumber\\
 & =: & N \left( p_F  \exp \left( c_x x_{T_1} + c_y y_{T_1} \right)  - \hat{K} \right)^+ , \label{twoasonepayoff1} 
\end{eqnarray}
for appropriate coefficients $p_F$, $c_x$, and $c_y$.

The expectation of (\ref{twoasonepayoff1}) under the $T_2$-forward measure is then equal to the undiscounted caplet price (to obtain the discounted caplet price, one would need to multiply by $P^D(0,T_{2})$).

If the day counts for caplet, forecasting forward rate, and discounting forward rate are different, the expressions for $\hat{K}$ and $p_F$ change correspondingly but the form of (\ref{twoasonepayoff1}) 
stays the same.

\subsection{Parametric Pricing setting}

We set up caplet pricing with given discount and forecasting curves for caplets for one fixed maturity and tenor, 
either over a range of strikes and/or over a range of model parameters. We used discount curve and forecasting curve 
as used for a particular day for a particular deal in USD. 
 
We use PDML and VML to learn the prices of these caplets as a function of contract and/or model parameters with uniform and adaptive parameter sampling. 
The script framework is used to generate samples of caplet prices and sample-wise derivatives with respect to chosen parameters for the training of 
the PDML and VML networks. 

We used the Cheyette-SV script in Figure \ref{cheyettescript1} in this paper.

\begin{figure}[h]  
\begin{lstlisting}
# function definition
g(x) = (1/mr)*(oneslike(x)-exp(-mr*x)) 

# system
d_ratevariance=vartheta*(1.0-positivepart(ratevariance))*d_t+\
               volofvar*ratevolatility*d_Z
ratevolatility=sqrt(positivepart(ratevariance))
d_ratex = (ratey-mr*ratex-g(measT-t)*ratevariance*volterm*volterm)*d_t+\
          ratevolatility*volterm*d_W 
d_ratey = (ratevariance*volterm*volterm-2.0*mr*ratey)*d_t
deltafwd = initfwd + hkd*(ratex+gkd*ratey)
volterm = volaterm*deltafwd+volbterm

# inital values
init: ratevariance = ones([batchsize])
init: ratex=zeros([batchsize])
init: ratey=zeros([batchsize])

# payoff
maturity: caplet pays positivepart(pf*exp(cx*ratex[fixingtime]+\
                                          cy*ratey[fixingtime])-khat) \
                 nodiscount 
                                          
\end{lstlisting}
\caption{Script for Caplet Pricing for Cheyette Model with SV (Euler for Cheyette, Euler Full Truncation for CIR) \label{cheyettescript1}} 
\end{figure}

\subsection{Pricing Results - Uniform Parameter Sampling}

We demonstrate the application of PDML with uniform parameter sampling for caplets in two settings. 
In the first setting, we vary a contract parameter (caplet strike) while keeping the model parameters fixed 
and learn a network that computes 
caplet price as a function of strike, for both VML and PDML. 
We regularize with sample-wise derivatives with respect to strike and also 
learn the derivative with respect to strike for PDML. 
In the second setting, we fix the contract parameter strike to ATM but we vary
two model parameters $a$ and $b$ and learn a network that computes ATM caplet price
as a function of those two model parameters, for both VML and PDML.
We regularize with sample-wise derivatives with respect to both model parameters and 
also learn the derivatives with respect to both for PDML. 
In both settings, we price caplets with one year maturity and 3 month (0.25 year) tenor.

{\bf Setting 1:} Here we are using fixed time-constant model parameters 
$a(t) \equiv -0.15873, b(t) \equiv 0.00788, \eta(t) \equiv 0.54224$. 
The parameter vector consists of only the contract parameter caplet strike. 
We generate samples $X_{i} = (K) \sim U(0.01,0.04)$. 
The prices and greeks obtained using 10M MC samples are considered as ground truth.  

\begin{figure}[h]
\centering
\subfigure[]{\includegraphics[width=0.9\linewidth]{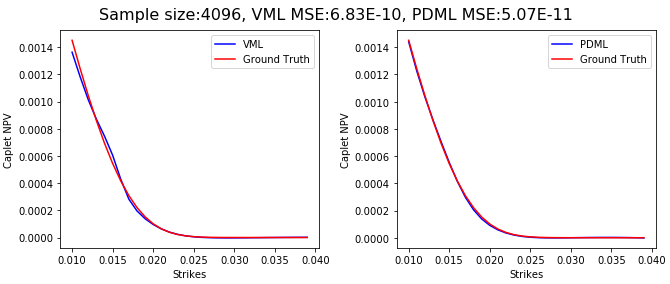}}
\subfigure[]{\includegraphics[width=0.9\linewidth]{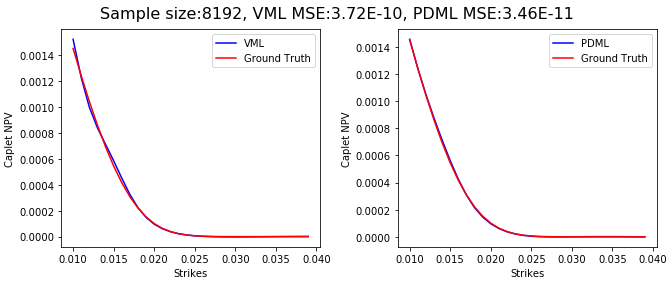}}
\subfigure[]{\includegraphics[width=0.9\linewidth]{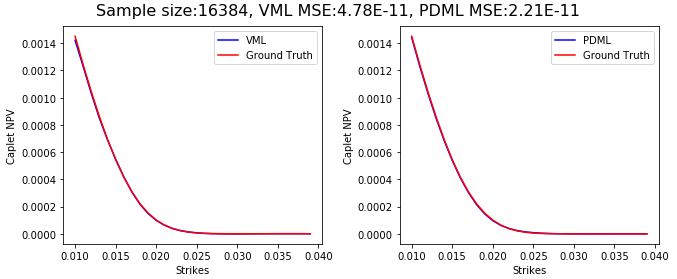}}
\caption{Comparison of Caplet Prices from VML vs PDML Across Various Sample Sizes}
\label{case1pricefig}
\end{figure}

In Figure \ref{case1pricefig}, we can see that PDML prices caplets more accurately than VML for a given sample size. 
In terms of sample efficiency, VML took $\sim 16K$ samples to achieve comparable accuracy obtained by PDML with $\sim 4K$ samples, 
thus PDML is about four times more sample efficient than VML.
 
\begin{figure}[h]
\centering
\subfigure[]{\includegraphics[width=0.9\linewidth]{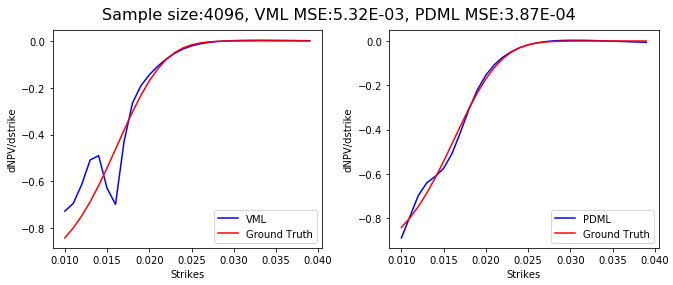}}
\subfigure[]{\includegraphics[width=0.9\linewidth]{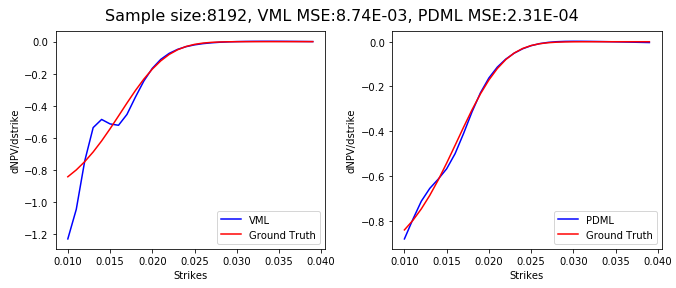}}
\subfigure[]{\includegraphics[width=0.9\linewidth]{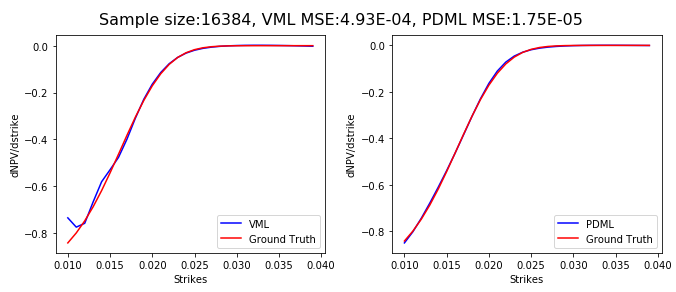}}
\caption{Comparison of Derivative of Caplet Price wrt Strike from VML vs PDML Across Various Sample Sizes}
\label{case1greeksfig}
\end{figure}

\begin{figure}[h]
\centering
\subfigure[]{\includegraphics[width=\linewidth]{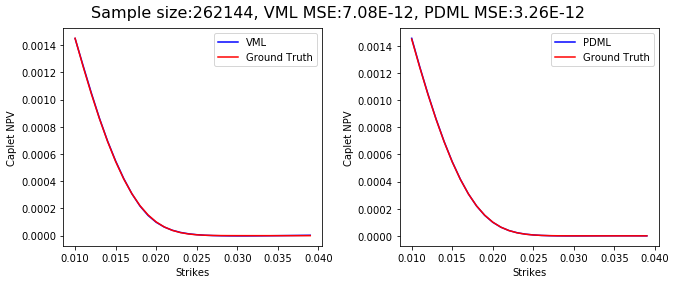}}
\subfigure[]{\includegraphics[width=\linewidth]{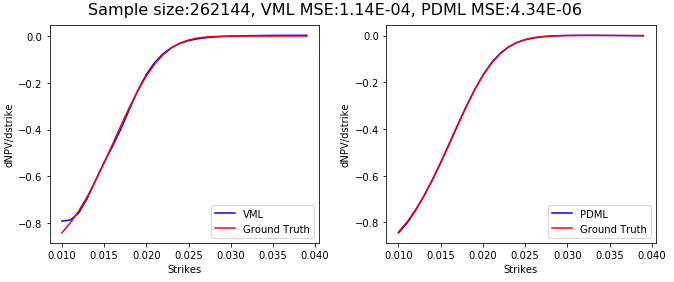}}
\caption{Prediction of Price and Derivative by VML and DML with Large Sample Sizes}
\label{case1largesamplefig}
\end{figure}

In Figure \ref{case1greeksfig}, we can see that for a given sample size, PDML predict 
derivatives with respect to strike better than VML. 
Furthermore, if the sample size is increased, the derivative with respect to strike
as computed by PDML approximates the ground truth very closely whereas VML only 
approximates the price well but not the derivative with respect to strike.
This can be seen from Figure \ref{case1largesamplefig} which uses large MC sample size $\sim 250K$. Both VML and PDML produce caplet prices close to ground truth. 
For the derivative with respect to strike, PDML  gives values close to ground truth while the VML results differ 
appreciably from the ground truth for small caplet strikes. 

Thus, from the numerical results observed in this setting, we can conclude that regularizing the training 
with differentials with respect to the varied contract parameter yield faster convergence to ground 
truth for both the price and the derivative of the price with respect to the contract parameter.

{\bf Setting 2:} In this setting, we consider varying model parameters and we fix the contract parameter (caplet strike) to ATM. 
We are using time-constant model parameters $a(t) \equiv a, b(t) \equiv b, \eta(t) \equiv 0.54224$. We 
generate varying model parameters  $X_{i} = (a,b) \sim U(-0.15,-0.18) \times U(0.0065,0.0085)$. 
We use $2^{16}$ MC samples  to train PDML \& VML networks and compare caplet prices and 
derivatives of them with respect to model parameters computed with those networks against 
ground truth prices and derivatives computed with $10M$ MC samples for each parameter set. 
    
\begin{figure}
\centering
\subfigure[]{\includegraphics[width=0.40\textwidth]{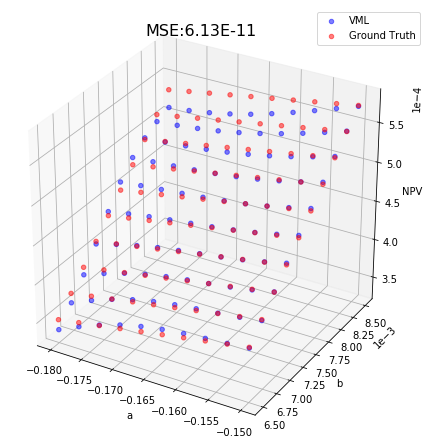}} 
\subfigure[]{\includegraphics[width=0.40\textwidth]{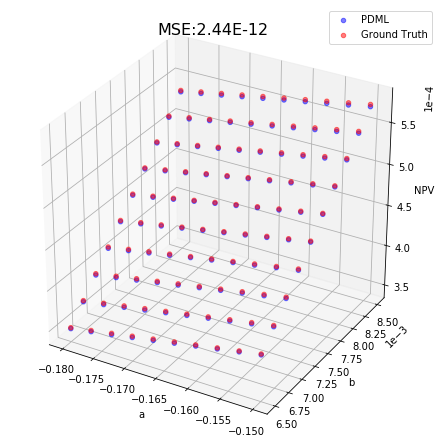}} 
\subfigure[]{\includegraphics[width=0.40\textwidth]{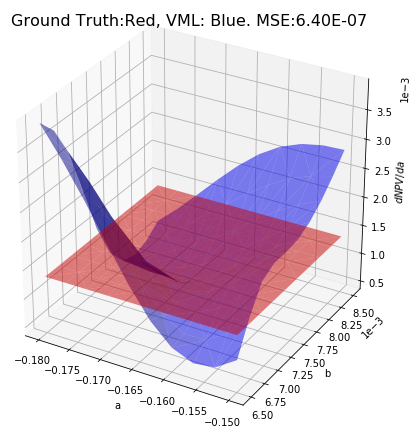}}
\subfigure[]{\includegraphics[width=0.40\textwidth]{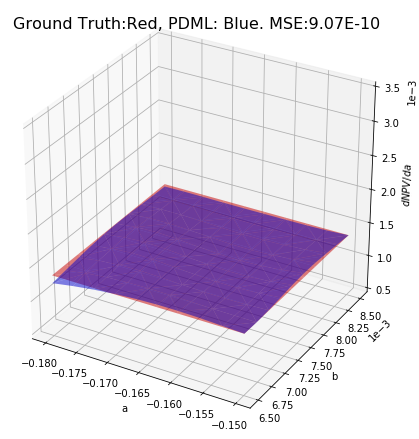}}
\subfigure[]{\includegraphics[width=0.40\textwidth]{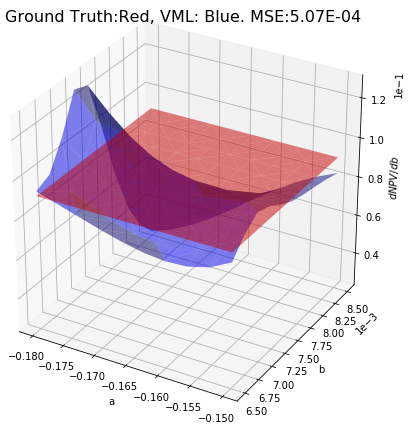}}
\subfigure[]{\includegraphics[width=0.40\textwidth]{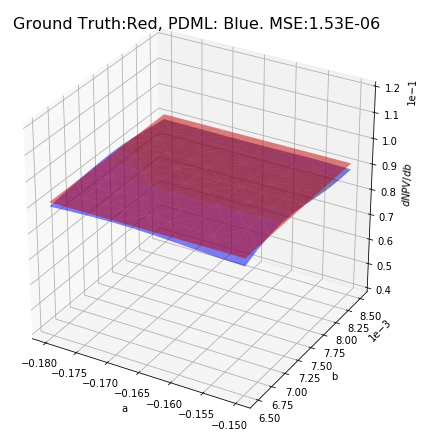}}
\caption{Comparison of PDML vs VML for Prices and for Risk Sensitivities wrt Model Parameters}
\label{case2resfig}
\end{figure}

In Figure \ref{case2resfig}, we can see that PDML network is more accurate than VML network in 
computing caplet derivative price. We also see that  VML network approximates risk sensitivities 
$\left(\frac{dNPV}{da}, \frac{dNPV}{db} \right)$ very poorly while PDML network 
approximates the risk sensitivities quite well for this given sample size.

\FloatBarrier

\subsection{Pricing Results - Adaptive Parameter Sampling}

\begin{figure}
\centering
\subfigure[]{\includegraphics[width=0.5\linewidth]{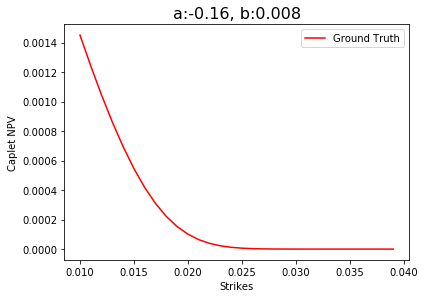}}
\subfigure[]{\includegraphics[width=0.5\linewidth]{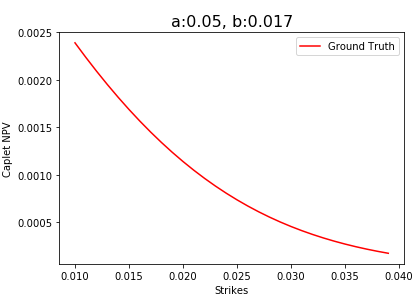}}
\subfigure[]{\includegraphics[width=0.5\linewidth]{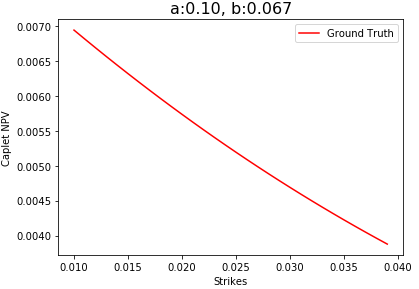}}
\caption{Ground Truth Prices for Different Parameter Sets Viewed Across Strikes. MC Prices with 10M Simulations are Considered as Ground Truth Prices.}
\label{diffpricemagnitudes}
\end{figure}

In Figure \ref{diffpricemagnitudes}, we show that 
different model parameter values (here three example parameter sets on model parameter space 
$\left(a(t),b(t)\right) \equiv \left(a,b\right) \equiv [-0.16,0.1] \times [0.008,0.067]$)
lead to conditional expectations (caplet prices) of different magnitude
across strike range $[0.01,0.04]$, in particular for different $b$. 

In calibration and other parametric pricing settings, it is important to achieve a certain relative accuracy 
for model parameters across the parameter range. 

\begin{figure}
\centering
\subfigure[]{\includegraphics[width=0.49\textwidth]{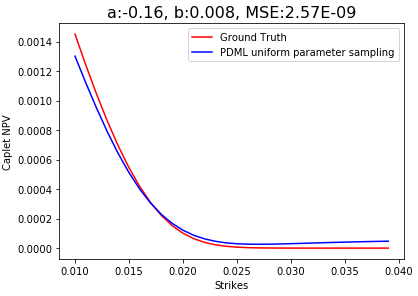}} 
\subfigure[]{\includegraphics[width=0.49\textwidth]{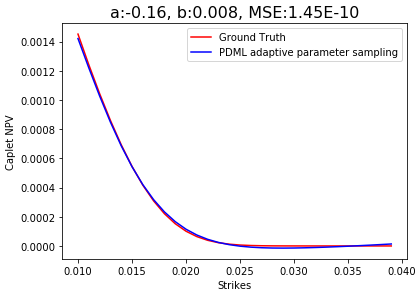}} 
\subfigure[]{\includegraphics[width=0.49\textwidth]{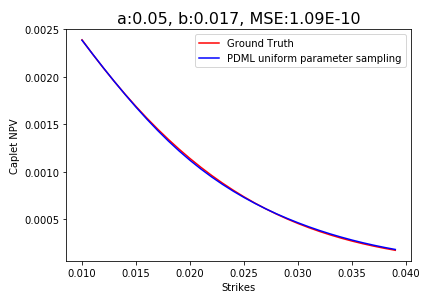}}
\subfigure[]{\includegraphics[width=0.49\textwidth]{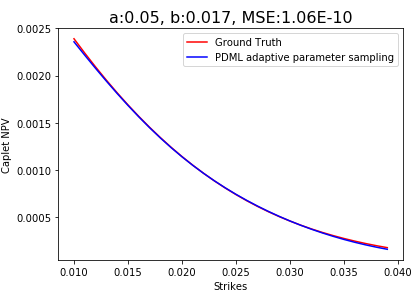}}
\subfigure[]{\includegraphics[width=0.49\textwidth]{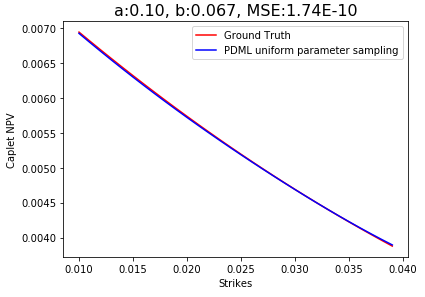}}
\subfigure[]{\includegraphics[width=0.49\textwidth]{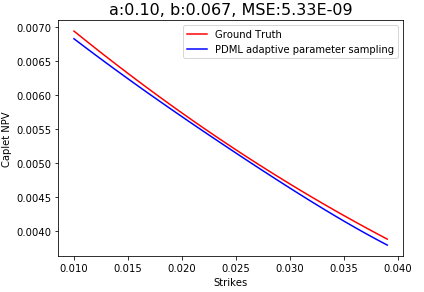}}
\caption{Comparison of PDML Approximation Accuracy with Uniform vs Adaptive
Parameter Sampling. Both PDML Surrogates Were Trained on MC Sample Sizes of
$2^{16}$.}
\label{uniformandadaptivesamplingfigs}
\end{figure}

In Figure \ref{uniformandadaptivesamplingfigs}, we compare 
PDML network trained under uniform parameter sampling (
$X = (a,b,k) \sim U[-0.16,0.1] \times U[0.008,0.067] \times U[0.01,0.04]$) 
on the left 
with a 
PDML network trained under adaptive parameter sampling (
$X = (a,b,k) \sim U[-0.16,0.1] \times P[0.008,0.067] \times U[0.01,0.04]$)
on the right. 
We can see on the left that for the top parameter set which results in conditional expectations smaller in magnitude, 
the PDML surrogate network does not approximate the ground truth well, while it does so in the lower two 
parameter sets for which the conditional expectations is larger in magnitude. 

We can see on the right that using an adaptive sampling $P[0.008,0.067]$ for the $b$ parameter leads to better approximation 
for the top two parameter sets and to a slightly worse approximation which mostly preserves shape. 

The adaptive sampling distribution P for parameter
b is such that $P(b) \propto \frac{1}{E\left[.|b \right]}$. These marginal
expectations $E\left[.|b \right]$ can be approximated by binning the samples
by the $b$ parameter and computing bin averages. Using these averages, 
we compute a cubic spline fit to get a smooth
approximation of $\frac{1}{E\left[.|b \right]}$. Finally, we scale the cubic
spline fit appropriately so that it integrates to one and can be used as
sampling distribution.  
 
In this way, we sample parameter sets leading to smaller conditional expectations more often so that 
we will approximate those parameters regions better than with an uniform distribution. This will also lead to 
fewer samples in those parameter regions leading to larger conditional expectations, leading to some degradation
of approximation quality there. If we want to approximate all regions as well as before, we will need to learn on 
a somewhat increased number of samples.

If we sample parameter ranges adaptively, MSE in regions with smaller magnitude of conditional expectation
will decrease and in regions with higher magnitude of conditional expectation, MSE will increase. 
But, in a relative sense, adaptive parameter sampling provides better results than 
uniform sampling distribution, as $\left(MSE_{adaptive}/MSE_{uniform}\right)$ 
increases by a factor of about 10 in the lower magnitude region
while it decreases by only $1/2$ in higher magnitude region.

\FloatBarrier

\section{Calibration Based on Parametric Pricing DNN Surrogates}\label{calibrationPDML}

Calibration is a very important step in many quantitative finance applications and a key step in pricing and risk management of vanilla and exotic options. 
In calibration, one tries to find a set of model parameters that fits the market data for calibration instruments best in a certain way. There might be additional 
constraints and regularizations beyond best fit to calibration instrument market data. Typically what is considered best is expressed as an objective function 
parametrized over model parameters that compares pricing under the model with given model parameters against prices given by market data. Least squares or weighted least 
squares is a common form of objective function. 

\subsection{Calibration by Optimization}

Calibration by optimization tries to find parameter sets that minimize that given objective function with appropriate optimization methods starting from appropriate initial 
guessses. Objective functions might have several local minima (which would be found by local optimization methods) and often one tries to find the global minimum or 
a good enough local minimum, thus favoring global or globalized optimizers. Typical calibration processes by optimization require calibration instruments repriced for
each term in the objective function. The calibration through optimization is typically run until some stopping criterion is satisfied which could be a certain level of objective 
function to achieve, a certain maximum number of optimization steps or generations to perform, or to stop once the parameters change less than a certain amount as measured in 
some metric or the objective function changes less than a certain number over the last step and/or generation, or a combination of such criteria. We assume that the optimization 
method identifies and returns one parameter set as the ``best" result, and we
will call that parameter set the optimized parameter set.

The standard approach generally involves the development of analytical or semi-analytical pricers for a simplified model 
and/or analytical or semi-analytical approximative pricers for the original problem. This means a major theoretical and implementation effort to first identify appropriate
simplifications, approximations, and assumptions and then implement both the simplification and approximation process and the (approximate) pricer for the (approximate) problem. 
These simplifications, approximations, and assumptions often introduce errors and biases so that close-to-optimal parameter sets for the approximate set-up are no longer 
close-to-optimal for the original model. 

Alternatively, one could use a possibly parallelized MC pricer for the vanilla calibration instruments insider of the global optimizer
(such as differential evolution - DE).  While this avoids the need of deriving approximations and simplifications and implementing special solvers, it is typically
very compute intensive and the number of MC paths or samples used in the pricer has to be kept small so that the calibration process can be run with the available computing
power. Often, this requires finding and implementing sufficiently good control variates or other variance reduction techniques, forcing the derivation and implementation 
of good enough control variates or the introduction of uncontrollable and unmonitored bias and/or variance due to the use of approximate MC sampling on a small number of
paths. 

We propose the use of PDML techniques 
to obtain one or several DNN surrogates for the pricing function derived from differentiable simulation 
  of the underlying model with apppropriately sampled parameters, 
to optimize with global optimizers (and potentially to refine with local optimizers) over these 
  surrogates, and 
to return a parameter set that performs well enough in a ground truth or reference pricer.

If calibrating to calibration instruments of different maturities, one can either proceed with a time-global calibration---where model parameters across all time intervals
are calibrated at the same time, allowing time-homogenous models or models with almost any kind of term structure---or a piecewise calibration, where constant model 
parameters are determined for each time interval between maturities. Often piecewise calibration (``bootstrapping") is preferred. For multiple maturities, we will discuss and use piecewise calibration. 

Thus, the calibration approaches that we will be using are the following: 

\textbf{Calibration Approach: Single Maturity.} To calibrate the model parameters from given caplet prices for maturity $T$, 
we assume the model parameters $\Theta_S$ are constant over the time period $[0,T]$, i.e., $a(t) \equiv a, b(t) \equiv b,$ and 
$\eta(t) \equiv \eta$ for the Cheyette model. 
Next, we set the interval bounds for input space and generate the training data set 
$\left\{X_{i},Y_{i}\right\}_{i=1}^{m}$ respectively $\left\{X_{i},Y_{i},DY_{i} \right\}_{i=1}^{m}$
where $X_{i} = \left(a^{i},b^{i},\eta^{i},k^{i}\right)$ is sampled using adaptive parameter sampling, 
with model parameters\footnote{
As mentioned earlier, mean reversion $\kappa$ will be given, and is typically calibrated and/or chosen according to 
other considerations and/or instruments, see discussion in \cite{andersenpiterbarginterest}.}
 $\Theta^i_S= \left(a^{i},b^{i},\eta^{i}\right)$ and contract 
parameters $\Theta^i_F= \left(k^{i}\right)$.
The output $Y_{i}$ is obtained as one MC sample and here it denotes the pathwise payoff of the caplet 
with strike price $k^{i}$, maturity $T$, and appropriate tenor (0.25/3M with the appropriate daycount). 
The output $DY_{i}$ (full set of derivatives of $Y_{i}$ with respect to the model parameters and the contract parameter
(strike)) can be obtained by algorithmic differentiation from the computational graph for the MC sampling 
and is computed in TensorFlow as needed. 

We use PDML technique with this training data to learn a surrogate  
$\hat{f}\left(X\right) \approx E\left[ Y | X  \right]$ 
where $Y$ is the caplet payoff. 
Then, finally we use $\hat{f}$ inside a (global) optimization method (such as ICDE) where different calibration contracts
correspond to varying $\Theta_F$ and find the optimized model
parameters $\Theta_S$ under calibration (here, for the Cheyette model, $\Theta_S=\left(\underline{a},\underline{b},\underline{\eta}\right)$.

\textbf{Calibration Approach: Multiple Maturities.} We perform piecewise-time constant model parameter calibration
based on given caplet prices and/or caplet volatility surface of different maturities and appropriate 3M tenor.
Organize the maturities $\left\{ T_{i} \right\}_{i=1}^{N}$ so that $0 = T_{0} <  T_{1} < \dots < T_{N}$. 
We assume $a(t) \equiv a_{i}, b(t) \equiv b_{i},$ and $\eta(t) \equiv \eta_{i}$ for $t \in [T_{i-1},T_{i})$, $i=1, \dots, N$ for this 
Cheyette model calibration or $\Theta_S(t) \equiv \Theta_{S,i}$ in the general case. 
The calibration procedure is as follows: 
First, we calibrate the model parameters of the left most time interval $[0,T_{1})$  
as described in the calibration approach for single maturity and 
obtain optimized parameters $\underline{\Theta}_{S,1}=\underline{a}_{1},\underline{b}_{1},\underline{\eta}_{1}$. 
We then fix these model parameters $\underline{\Theta}_{S,1}=\underline{a}_{1},\underline{b}_{1},\underline{\eta}_{1}$ 
for time interval $[0,T_{1})$ and repeat for the next consecutive time interval $[T_{1},T_{2})$ 
to obtain optimized parameters $\underline{\Theta}_{S,2}=\underline{a}_{2},\underline{b}_{2},\underline{\eta}_{2}$. 
We repeat until we calibrate the model parameters of right most interval $[T_{N-1},T_{N}]$ and 
obtain $\underline{\Theta}_{S,N}=\underline{a}_{N},\underline{b}_{N},\underline{\eta}_{N}$.

\subsection{Global Optimization Methods}

To perform the optimization for the calibration, appropriate optimization methods are needed. 
We will at first use generic global or globalized optimization methods that can reliably handle 
optimization problems of various kinds, even though any particular optimization problem that 
we apply it to might be such that simpler or more specialized optimization methods might work. Even 
though PDML provides parameter derivatives, the chosen optimization method might not take (full)
advantage of it. While PDML surrogates are in general at least as accurate as VML and more accurate 
than MC pricers with small number of paths, they might be of limited and not specified accuracy,\footnote{We 
will discuss further below how we addressed this to a certain extent by training several different 
surrogates based on different random seeds.} 
and thus trying to solve the optimization problems more accurately than the surrogates are might 
not be advisable. At least some of the models typically calibrated in quantitative finance 
(such as Heston) are known to have involved topology and geometry and allow several local minima, 
thus we are lead to consider global or globalized optimization methods.

There are many global and globalized optimizers described in the literature. 
To name a few: for global approches, there are a variety of differential evolution (DE) variants 
and particle swarm optimization (PSO) variants, simulated annealing (SA), and other genetic algorithms (GA) or evolutionary algorithms (EA). 
Globalized approaches include the usual local optimization methods such as 
Gauss-Newton, Newton-Raphson, L-BFGS etc. but started from random or otherwise widely sampled parameter sets.
One can also start these local methods with populations or sets of parameter intervals 
determined by some first stage algorithm.

We had previously implemented an improved constrained differential evolution (ICDE) global optimization routine in Python and in C++ 
\cite{jia2013improved} which performs well and can be easily parallelized while also being able to handle 
equality and inequality constraints. Besides the equality and inequality constraints, one is also given 
a (hyper-)rectangular search space defined by intervals for each component of the parameter 
vector under minimization. ICDE is a further developed variant of differential evolution \cite{price2005differential}.
Differential evolution is a simple yet efficient evolutionary algorithm starting from some initial population 
that is typically randomly sampled from the given intervals; and then undergoes crossover, mutation, selection, 
enforcement of boundary constraints etc. to produce a number of generations. ICDE \cite{jia2013improved}
uses crossover strategies rand/1, rand/2, current to rand/1, and current to best/1 and improved breeder genetic 
algorithm (IBGA). Constraints are treated by minimizing the degree of constraint violation as defined appropriately.\footnote{
For details, we refer to \cite{jia2013improved}.}
All the parameter sets to be evaluated for the next generation are collected and then evaluated in parallel with several 
backends to allow parallel processing through Python's multiprocessing or joblib facilities or similar.

In our tests, this ICDE implementation performed well enough. In future work, we intend to test other global and/or globalized 
optimization methods that potentially could take advantage of derivative information, and possibly lead to faster and/or 
more robust methods. 

\subsection{Impact of Randomness on Surrogates: Making Optimization More Robust}

When calibrating with surrogates obtained by PDML, we are using random numbers at several points during the process. First, 
we consume random numbers during the parameter and MC sampling to generate $X$, $Y$, and $DY$.
Then, the DNN for PDML are initialized according to some random initializations (in our DNN architecture, 
these would be the weights and biases). 

Then, the DNN in the PDML step are trained by stochastic gradient descent methods with Adam Momentum, 
which might consume additional random numbers. 
Once the surrogates have been constructed, the global optimization method (such as ICDE) might also 
consume random numbers (in fact, ICDE uses random numbers to construct a new population of parameter sets 
from the old one). All these random numbers could be generated from either a single stream or from several 
streams with their own seed(s). Different choice of such seeds will in general lead to variation in results
and thus training several surrogates in parallel and optimizing against them will lead in general to 
different surrogates and different optimized parameter sets. These parameter sets associated to different
seeds might correspond to differently accurate DNN networks that lead to optimized parameter sets of 
different accuracies. 

Running independent constructions of PDML surrogates and optimizing over them starting with different 
seeds on different instances is perfectly parallel. Given the multi-core nature of common commodity 
CPU, one can obtain 5-10 replications at essentially the same running time as one replication; and on elastic compute 
units or larger server farms, this scales to even higher number of replications. 

Once we have constructed several replications in parallel in this way, we quantify the variation 
in the results of those replications  through two metrics: `PDML fit error' and `Model error' as follows:
\begin{eqnarray}\label{calibrationmetrics1}
\text{PDML fit error} &=& \sum_{\text{instruments}} \left(\text{PDMLapproximation} - \text{MC estimate} \right)^{2} \\
\text{Model error} &=& \sum_{\text{instruments}} \left(\text{MC estimate} - \text{target price}\right)^{2}
\end{eqnarray}  
Here, PDMLapproximation stands for the pricing with the PDML surrogate in that replication with the optimized parameter set from that replication,
MC estimate stands for the MC pricing with the optimized parameter set from that replication with enough samples (to estimate ground truth), 
and target price stands for the price of the calibration instruments as provided by market data or market data objects (such as caplet volatility 
surfaces together with appropriate Black or Bachelier formulas). 

For the $i^{th}$ step in the multi-maturity calibration procedure, these metrics can be re-written as follows:
\begin{eqnarray}\label{calibrationmetrics2}
\text{PDML fit error} &=& \sum_{\Theta_F} \left( \hat{f}\left(\underline{\Theta}_{S,i},\Theta_F \right)     
-  MC \left(\underline{\Theta}_{S,i},\Theta_F \right) \right)^{2} \\
\text{Model error} &=& \sum_{\Theta_F} \left( MC \left(\underline{\Theta}_{S,i},\Theta_F \right) -  MKT(\Theta_F) \right)^{2}
\end{eqnarray}  
where $\Theta_F=(k)$ denotes the contract parameter(s) (caplet strike in our example), 
$MC(.)$ denotes the MC price obtained by enough samples (to estimate ground
truth) with $\underline{\Theta}_{S,i}=\left(\underline{a}_{i},\underline{b}_{i},\underline{\eta}_{i} \right)$ as model parameters and 
$MKT(\Theta_F)$ denotes the market or target price.

Based on these metrics, we can select a best seed or set of best seeds to define some 
robustification to the optimization. That robustification will have results less dependent 
on any particular seed and likely perform better than an optimization run with only one particular seed. 
We propose two robust calibration approaches, namely, best seed approach and ensemble approach.    
\newline
\newline
The best seed approach is as follows:
\begin{enumerate}
\item Set seed for NN weight initialization, optimization algorithm, MC sample generation and run calibration. Repeat the process with several seeds.
\item Select the best seed based on two metrics as follows: 
\begin{equation}
\text{bestseed} = \min_{\text{seed}} \max \left(\text{PDML fit error}, \text{Model error} \right)
\end{equation}
\item The optimal model parameters corresponding to the best seed are selected.
\end{enumerate}
The ensemble approach is as follows:
\begin{enumerate}
\item Set seed for NN weight initialization, optimization algorithm, MC sample generation and run calibration. Repeat  the process with several seeds.
\item Pick a set of seeds with least $\max \left(\text{PDML fit error}, \text{Model error} \right)$.
\item Use the PDML networks corresponding to the selected seeds from the previous step to form an ensemble network. This ensemble network can be formed by averaging the results from individual component networks. 
\item Finally, run the calibration part again but use the ensemble network as surrogate and obtained optimized parameters. 
\end{enumerate}
We have observed that the seed for the global optimization in the calibration has a relatively small impact on the results 
and thus we pick the seed corresponding to the best replication in the final calibration against the ensemble.   

\section{Calibration Setting and Results}\label{calibrationresults}

\subsection{Calibration Setting}

We calibrate Cheyette models from caplet prices for maturities from 1 year to 6 years, all for tenor 3M, with the PDML technique. 
We show calibrations to single maturity data and settings and multi-maturity data and settings. We demonstrate both single-seed 
as well as robustified optimizations.

As is common when calibrating Cheyette models and other similar interest rate models, mean reversion speed $\kappa$ for the Cheyette SDEs and ODEs and
mean reversion speed $\theta$ for the stochastic volatility process are chosen according to other considerations and instruments before calibrating 
to caplets or swaptions (see \cite{andersenpiterbarginterest} for a discussion). In this calibration section, we fix them to the values $\kappa=0.03$ and $\theta=0.2$ 
for our numerical tests.

\subsection{Single Seed Calibration Approach: Single Maturity}
In single seed calibration approach, we use one single seed for NN weight initialization, 
DE optimization algorithm, and MC sample generation. In this section, we report 
numerical results when we calibrate models with constant model parameters against
single-maturity data corresponding to different maturities. 
In particular, we work with six yearly maturities starting from 1 year 
to 6 years. We repeat the calibration with two different seeds. 

\begin{figure}
    \centering
    \subfigure[]{\includegraphics[width=0.49\textwidth]{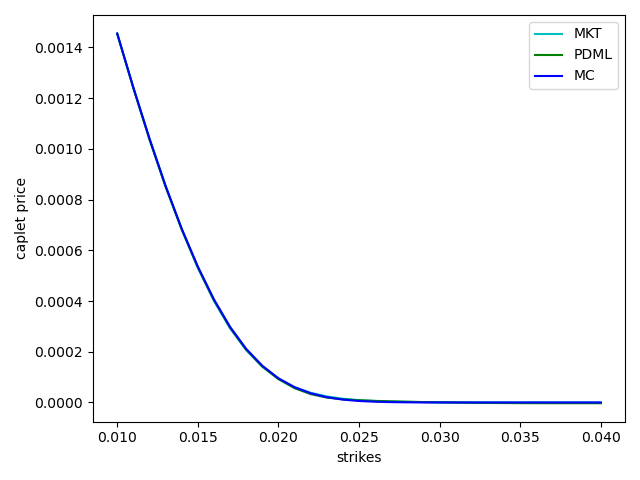}} 
    \subfigure[]{\includegraphics[width=0.49\textwidth]{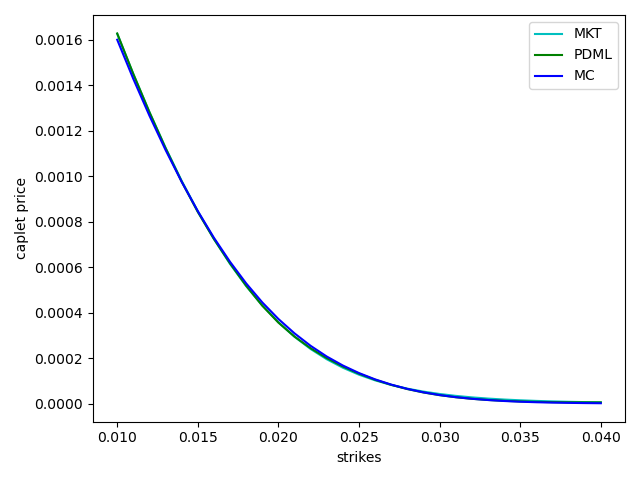}} 
    \subfigure[]{\includegraphics[width=0.49\textwidth]{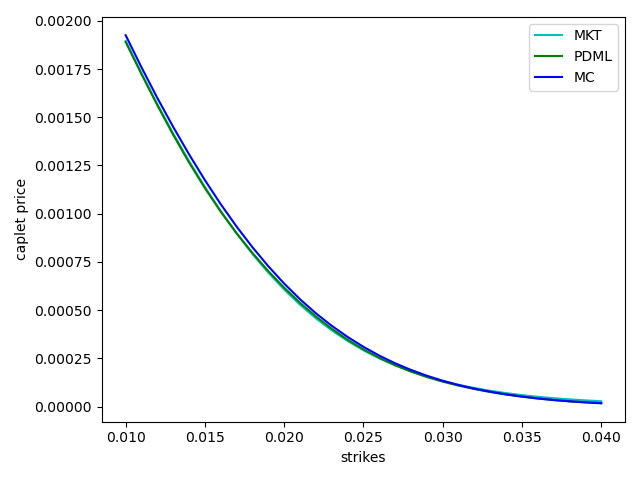}}
    \subfigure[]{\includegraphics[width=0.49\textwidth]{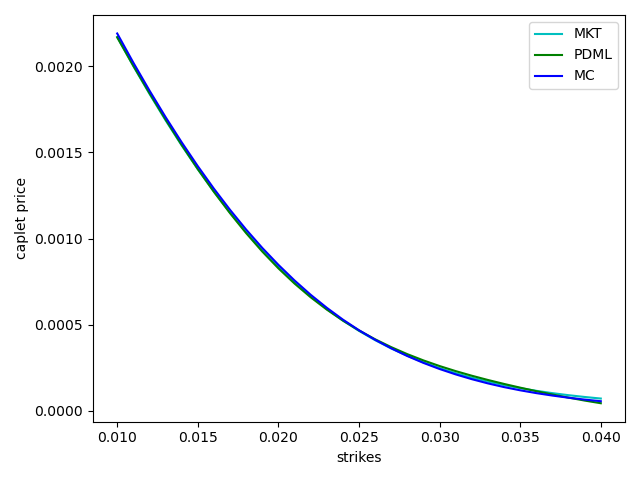}}
    \subfigure[]{\includegraphics[width=0.49\textwidth]{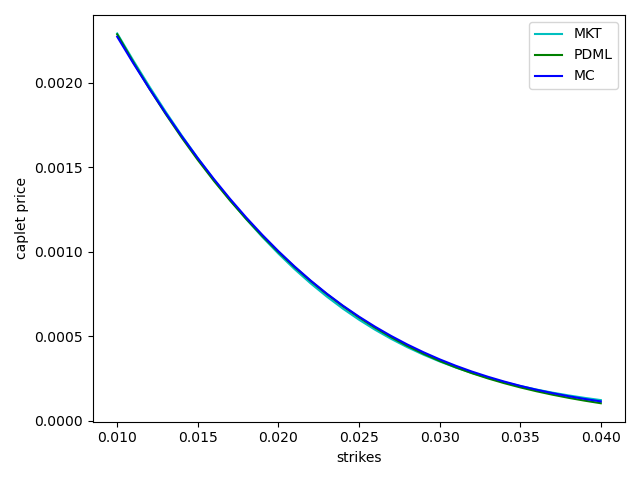}}
    \subfigure[]{\includegraphics[width=0.49\textwidth]{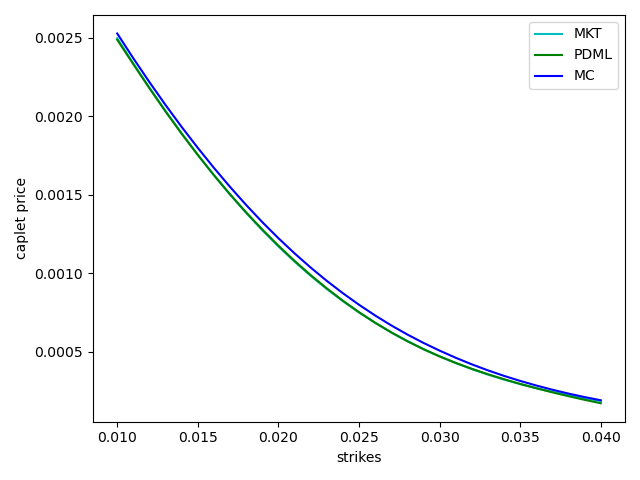}}
    \caption{Single Seed Calibration Approach: Single Maturity. Prices From PDML Network Trained Using Seed:2476 and MC for Optimized Parameter Sets Compared to 
    Market Prices. (a) 1yr Maturity (b) 2yr Maturity (c) 3yr Maturity (d) 4yr Maturity (e) 5yr Maturity (f) 6yr Maturity. }
    \label{price_graphs_singleseed_1}
\end{figure}

\begin{figure}
    \centering
    \subfigure[]{\includegraphics[width=0.49\textwidth]{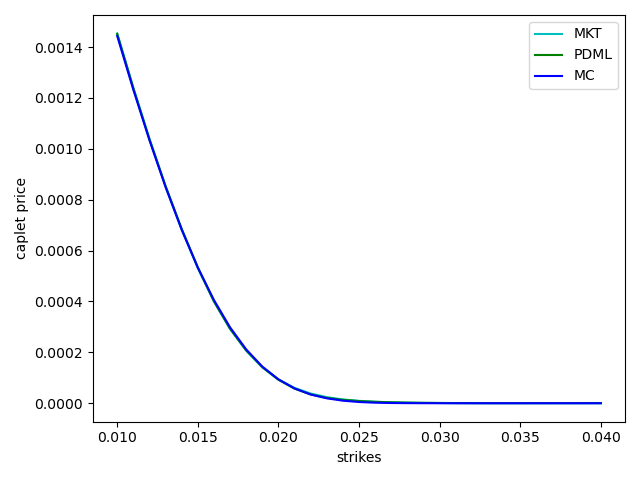}} 
    \subfigure[]{\includegraphics[width=0.49\textwidth]{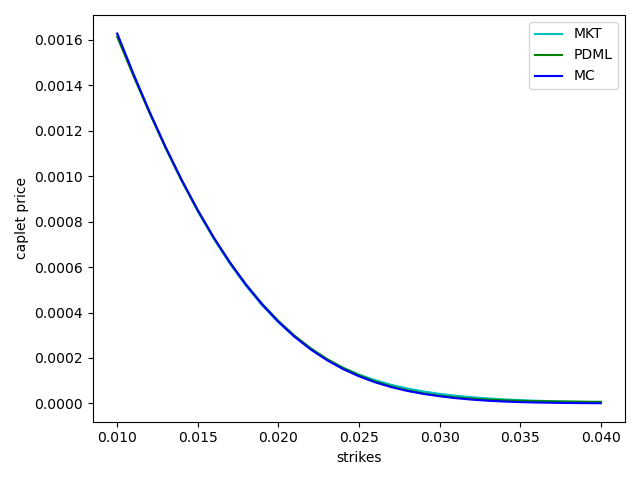}} 
    \subfigure[]{\includegraphics[width=0.49\textwidth]{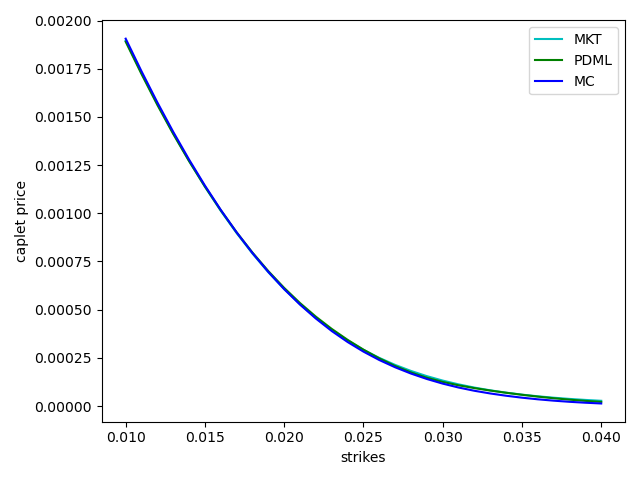}}
    \subfigure[]{\includegraphics[width=0.49\textwidth]{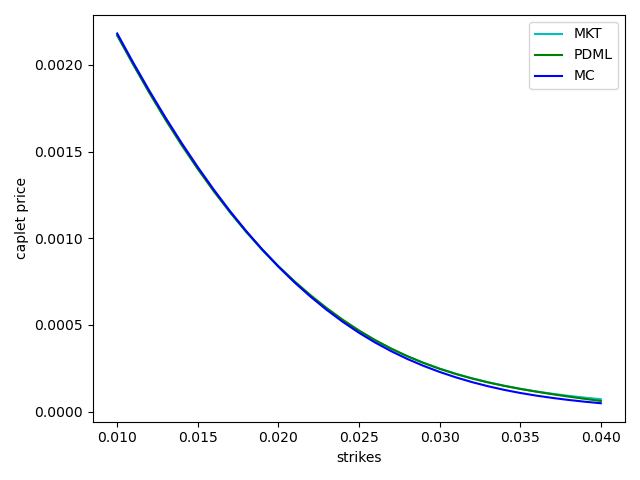}}
    \subfigure[]{\includegraphics[width=0.49\textwidth]{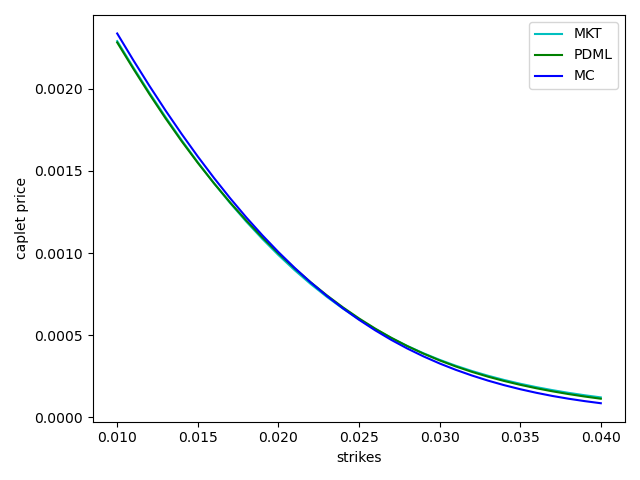}}
    \subfigure[]{\includegraphics[width=0.49\textwidth]{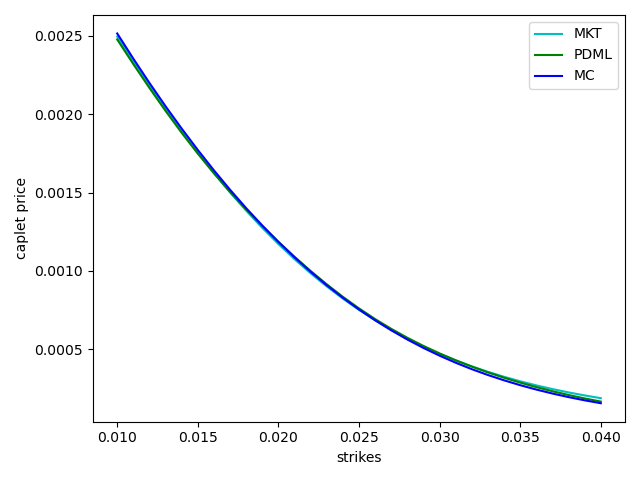}}
    \caption{Single Seed Calibration Approach: Single Maturity. Prices From PDML Network Trained Using Seed:5670 and MC at Optimized Parameter Sets Compared to  
    Market Prices. (a) 1yr Maturity (b) 2yr Maturity (c) 3yr Maturity (d) 4yr Maturity (e) 5yr Maturity (f) 6yr Maturity. }
    \label{price_graphs_singleseed_2}
\end{figure}

Figures \ref{price_graphs_singleseed_1} and \ref{price_graphs_singleseed_2} depict the results corresponding to the two different seeds. 
In both figures, results are reasonably accurate except 3 year and 6 year maturities in Figure \ref{price_graphs_singleseed_1} and 
except 4 year and 5 year maturities in Figure \ref{price_graphs_singleseed_1}. Thus, using any single seed may or may not yield 
optimized parameter sets with reasonable accuracy for all maturities. 
Therefore, we are interested in more robust calibration approaches that are less sensitive to particular seeds and give 
accurate enough results for tests at all maturities.  

\subsection{Single Seed Calibration Approach: Multiple Maturities}

In Table \ref{table:1},  we consider five different set of seeds and for each seed we report the two metrics
when calibrating against multiple maturities. 
We can see that using one single seed may or may not lead to uniformly good or comparable accuracy for all maturities. 
If by chance that single seed is ``bad" for some or most maturities, results
will be less accurate and calibration will be less reliable.

\begin{table}[h!]
\centering
\begin{tabular}{ |c|c|c|c|c|c|c| }
\hline
Maturity & Metric & Seed:2476 & Seed:4548 & Seed:5670 & Seed:5818 & Seed:8642 \\
\hline
\multirow{2}{4em}{1 yr} & PDML fit error & 9.13E-11 & 6.49E-12  & 8.71E-12  & 1.11E-11  & 2.40E-12 \\
& Model error & 8.90E-11  & 9.53E-12  & 1.42E-11 & 1.92E-11  & 9.62E-12 \\
\hline
\multirow{2}{4em}{2 yr} & PDML fit error & 5.14E-10   & 3.18E-11  & 2.34E-11  & 8.62E-11 & 2.20E-11 \\
& Model error & 5.17E-10 &  2.91E-11 & 4.62E-11 & 1.15E-10 & 3.43E-11 \\
\hline
\multirow{2}{4em}{3 yr} & PDML fit error & 3.03E-10  & 1.92E-11  & 1.65E-11   & 2.49E-10  & 8.98E-12  \\
& Model error & 2.96E-10 & 3.32E-11 & 6.52E-11 & 2.83E-10 &  1.81E-11 \\
\hline
\multirow{2}{4em}{4 yr} & PDML fit error & 1.02E-09  & 5.83E-11  & 3.47E-11  & 2.78E-11  & 8.54E-11 \\
& Model error & 1.02E-09 & 7.02E-11  & 1.14E-10  & 1.10E-10 & 8.65E-11\\
\hline
\multirow{2}{4em}{5 yr} & PDML fit error & 1.24E-09 & 1.76E-10   & 9.42E-11  & 2.48E-10 & 8.97E-11 \\
& Model error & 1.23E-09 & 2.12E-10  & 2.09E-10 & 3.33E-10 & 8.21E-11\\
\hline
\multirow{2}{4em}{6 yr} & PDML fit error & 1.61E-09  & 1.46E-10  & 1.34E-10  & 3.87E-10   & 5.62E-11 \\
& Model error & 1.62E-09 & 1.99E-10  & 3.03E-10 & 5.00E-10 &  8.17E-11 \\
\hline
\end{tabular}
\caption{Single Seed Calibration Approach: Multiple Maturities. Calibration Results for Multiple Maturities Calibrations for Five Different Seeds.}
\label{table:1}
\end{table}

\subsection{More Robust Calibration Approach: Single Maturity}

In this section, we test the more robust calibration approaches that we proposed for the single maturity case.
In both of these robust calibration approaches, we run the calibration with five different seeds. 
For the best seed approach, we pick the best seed. For the ensemble, we form an ensemble network based on the 
best three seeds. 

\begin{figure}
    \centering
    \subfigure[]{\includegraphics[width=0.49\textwidth]{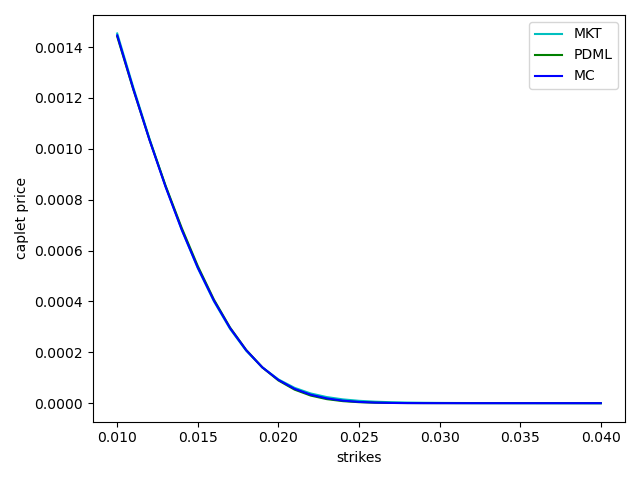}} 
    \subfigure[]{\includegraphics[width=0.49\textwidth]{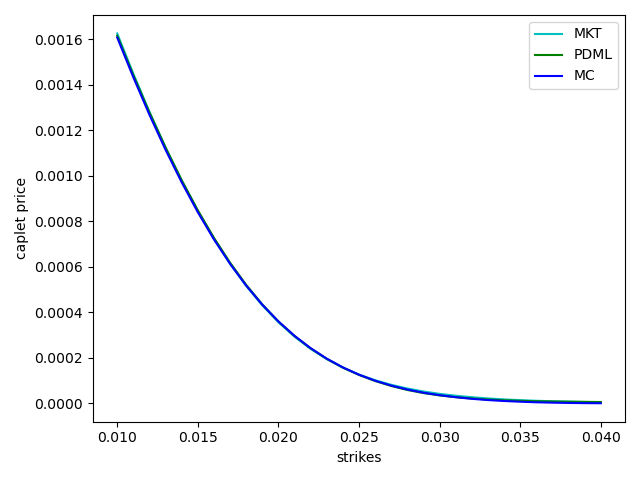}} 
    \subfigure[]{\includegraphics[width=0.49\textwidth]{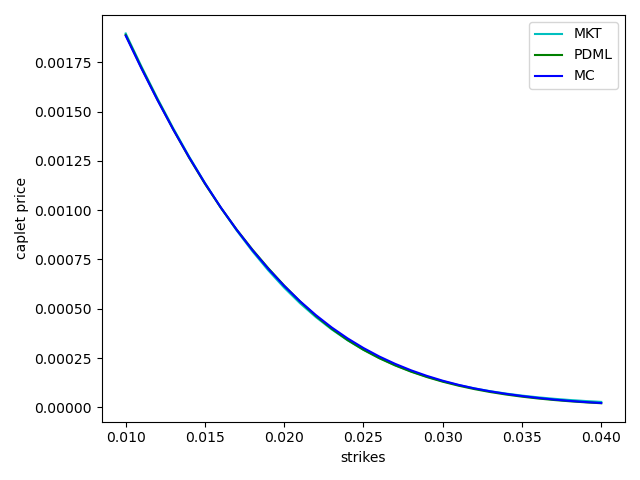}}
    \subfigure[]{\includegraphics[width=0.49\textwidth]{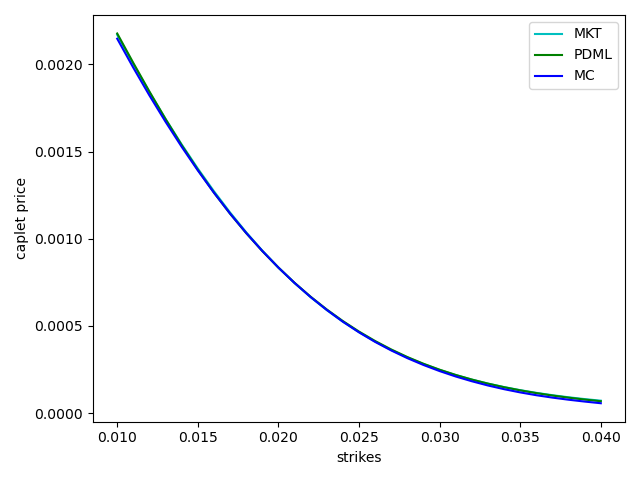}}
    \subfigure[]{\includegraphics[width=0.49\textwidth]{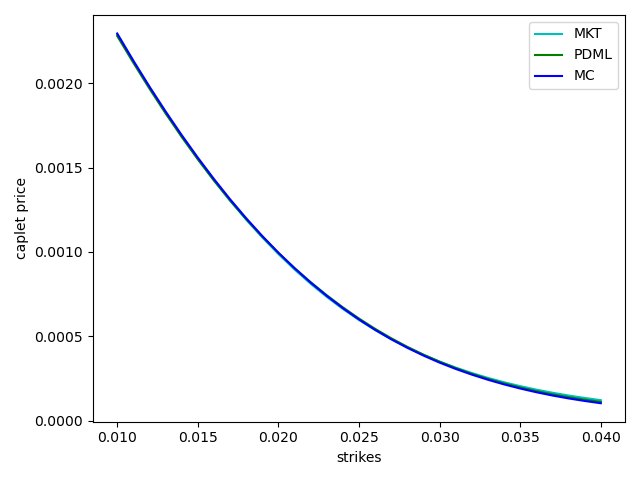}}
    \subfigure[]{\includegraphics[width=0.49\textwidth]{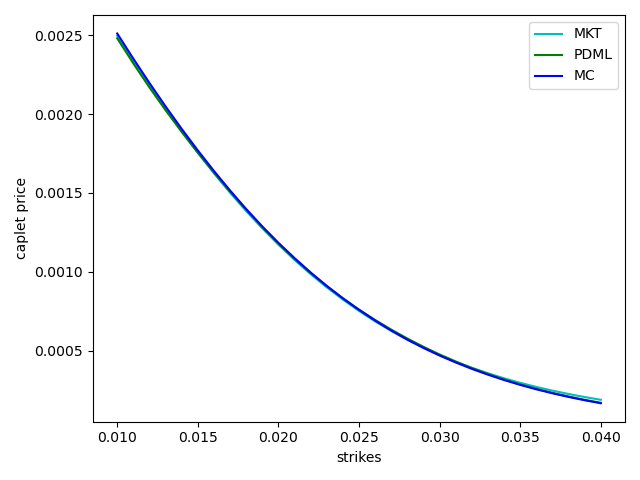}}
    \caption{More Robust Calibration Approach: Single Maturity. Prices From Best Seed PDML Network and MC at Optimized Parameter Sets Compared to Market Prices. (a) 1yr Maturity (b) 2yr Maturity (c) 3yr Maturity (d) 4yr Maturity (e) 5yr Maturity (f) 6yr Maturity. }
    \label{price_graphs_singlematurity_bestseed}
\end{figure}

\begin{figure}
    \centering
    \subfigure[]{\includegraphics[width=0.49\textwidth]{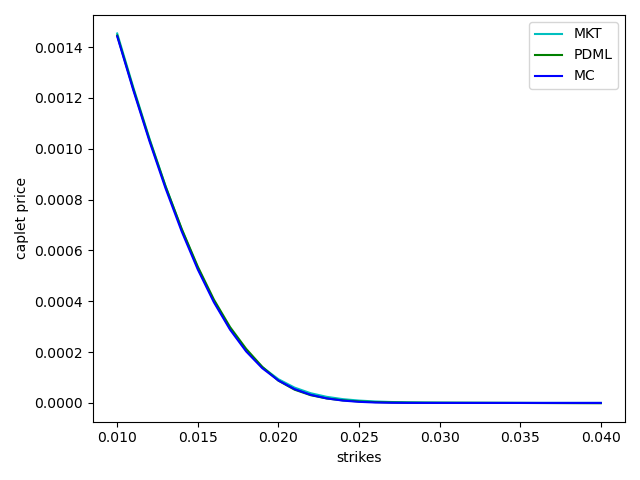}} 
    \subfigure[]{\includegraphics[width=0.49\textwidth]{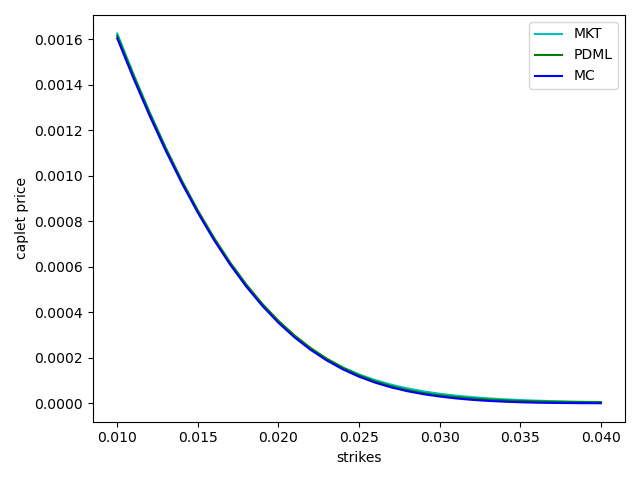}} 
    \subfigure[]{\includegraphics[width=0.49\textwidth]{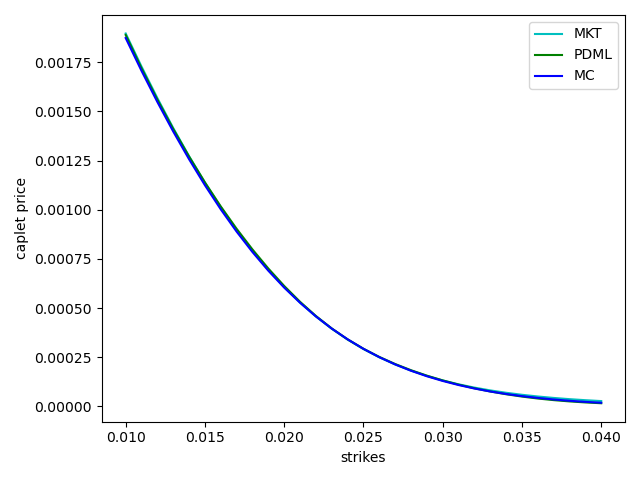}}
    \subfigure[]{\includegraphics[width=0.49\textwidth]{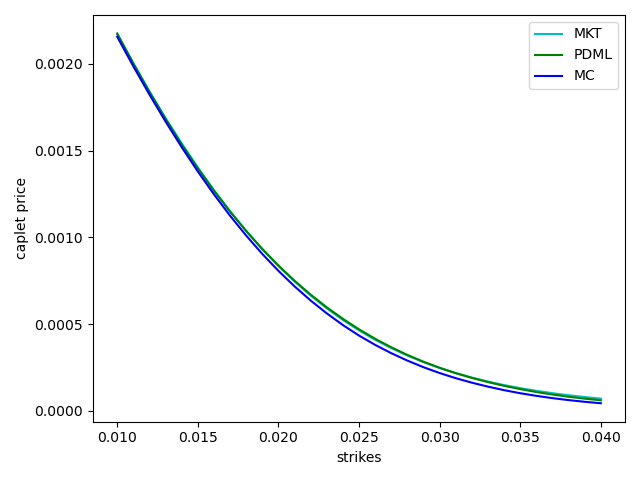}}
    \subfigure[]{\includegraphics[width=0.49\textwidth]{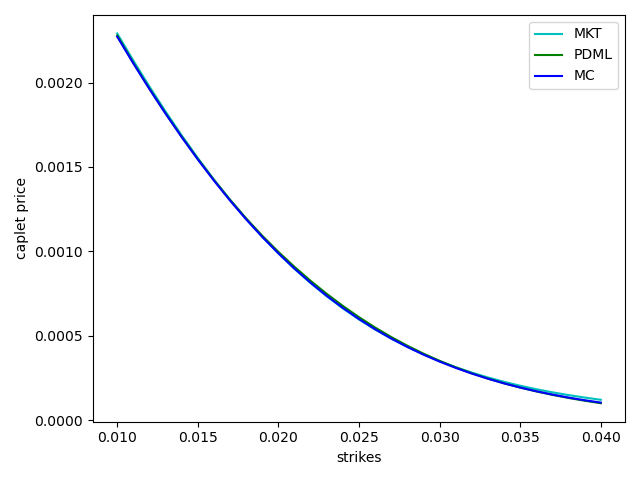}}
    \subfigure[]{\includegraphics[width=0.49\textwidth]{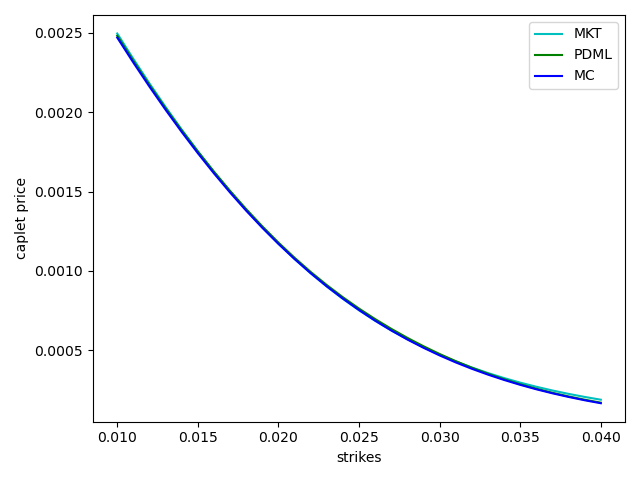}}
    \caption{More Robust Calibration Approach: Single Maturity. Prices From Ensemble PDML Network and MC at Optimized Parameter Sets Compared to Market Prices. (a) 1yr Maturity (b) 2yr Maturity (c) 3yr Maturity (d) 4yr Maturity (e) 5yr Maturity (f) 6yr Maturity. }
    \label{price_graphs_singlematurity_ensemble}
\end{figure}

In Figure \ref{price_graphs_singlematurity_bestseed},
we see that the best seed calibration approach yielded sufficiently accurate results for all the maturities considered. 
In Figure \ref{price_graphs_singlematurity_ensemble}, 
we present the results corresponding to an ensemble PDML network corresponding to an ensemble made from the PDML networks
corresponding to the best three seeds. 
Except for 4 year maturity, all maturities are calibrated sufficiently accurate with the ensemble approach also. 
In further testing, we observed that for the 4 year maturity case, 
the best two seeds produced quite accurate networks while the third best 
performed quite a bit worse in terms of metrics and affected the quality of the ensemble network. 
One could either consider some adaptive selection of seeds 
or form ensembles starting from more seeds and selecting the best three seeds from those 
(say, ensemble best 2 out of 5 or best 3 out of 8). In this particular 
case, a best 2 out of 5 ensemble would have given sufficiently accurate results for all maturities.

\subsection{More Robust Calibration Approach: Multiple Maturities}
In Table \ref{table:2}, we report results from the best seed approach. 
The best seed is chosen such that it has least max error 
(highlighted in bold). We can see that the best seed approach 
resulted in lower PDML fit error and Model error across all maturities. 
\begin{table}[h!]
\centering
\begin{tabular}{ |c|c|c|c|c|c|c| }
\hline
Maturity & Metric & Seed:2476 & Seed:4548 & Seed:5670 & Seed:5818 & Seed:8642 \\
\hline
\multirow{3}{4em}{1 yr} & PDML fit error & 9.13E-11 & 6.49E-12 & 8.71E-12 & 1.11E-11  & 2.40E-12 \\
& Model error & 8.90E-11  & 9.53E-12 & 1.42E-11  & 1.92E-11  & 9.62E-12 \\
&max error & 9.13E-11 & \textbf{9.53E-12}  & 1.42E-11  & 1.92E-11  & 9.62E-12 \\
\hline
\multirow{3}{4em}{2 yr} & PDML fit error & 1.81E-10 & 2.30E-11 & 1.17E-11  & 8.40E-12 & 8.77E-11 \\
& Model error & 1.84E-10  & 3.17E-11  & 2.47E-11 & 2.78E-11  & 9.73E-11 \\
&max error & 1.84E-10  & 3.17E-11  & \textbf{2.47E-11} & 2.78E-11 & 9.73E-11 \\
\hline
\multirow{3}{4em}{3 yr} & PDML fit error & 9.69E-10 & 1.82E-11 & 1.26E-10  & 2.44E-11  & 6.72E-11 \\
& Model error & 9.73E-10 & 2.80E-11  & 1.27E-10  & 2.38E-11  & 8.92E-11 \\
&max error & 9.73E-10  & 2.80E-11  & 1.27E-10  & \textbf{2.44E-11}  & 8.92E-11 \\
\hline
\multirow{3}{4em}{4 yr} & PDML fit error & 1.44E-10 & 2.01E-11  & 2.39E-10  & 4.40E-11  & 3.24E-10 \\
& Model error & 1.21E-10  & 3.79E-11 & 2.62E-10  & 6.00E-11  & 3.16E-10 \\
&max error & 1.44E-10 & \textbf{3.79E-11}  & 2.62E-10 & 6.00E-11 &  3.24E-10 \\
\hline
\multirow{3}{4em}{5 yr} & PDML fit error & 1.02E-09 & 3.46E-12 & 6.20E-11  & 3.22E-11  & 2.35E-10 \\
& Model error & 1.02E-09 & 4.09E-11  & 6.20E-11  & 6.37E-11  & 2.25E-10 \\
&max error & 1.02E-09 & \textbf{4.09E-11}  & 6.20E-11  & 6.37E-11  & 2.35E-10 \\
\hline
\multirow{3}{4em}{6 yr} & PDML fit error & 2.14E-09 & 1.35E-10 & 2.27E-10  & 4.32E-11  & 1.26E-10 \\
& Model error & 2.12E-09  & 1.90E-10  & 2.65E-10  & 1.09E-10  & 1.49E-10 \\
&max error & 2.14E-09 &  1.90E-10  & 2.65E-10  & \textbf{1.09E-10}  & 1.49E-10 \\
\hline
\end{tabular}
\caption{More Robust Calibration Approach: Multiple Maturities. Calibration Results for Multiple Maturities From Best Seed Approach.}
\label{table:2}
\end{table}

\begin{figure}
    \centering
    \subfigure[]{\includegraphics[width=0.49\textwidth]{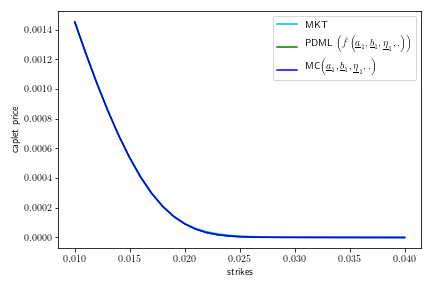}} 
    \subfigure[]{\includegraphics[width=0.49\textwidth]{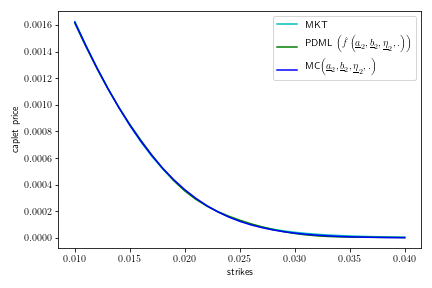}} 
    \subfigure[]{\includegraphics[width=0.49\textwidth]{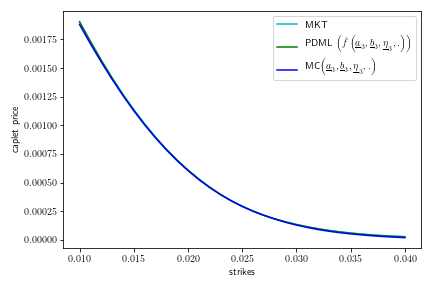}}
    \subfigure[]{\includegraphics[width=0.49\textwidth]{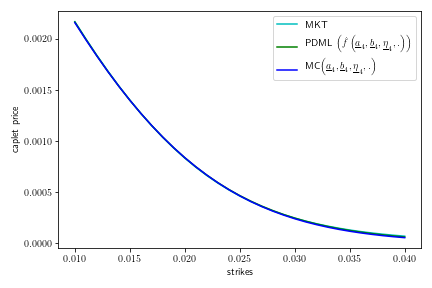}}
    \subfigure[]{\includegraphics[width=0.49\textwidth]{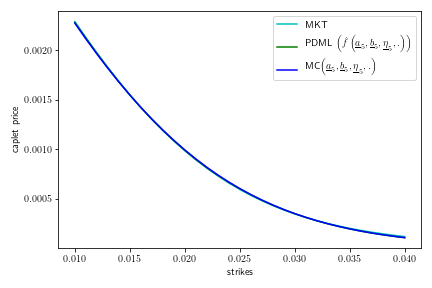}}
    \subfigure[]{\includegraphics[width=0.49\textwidth]{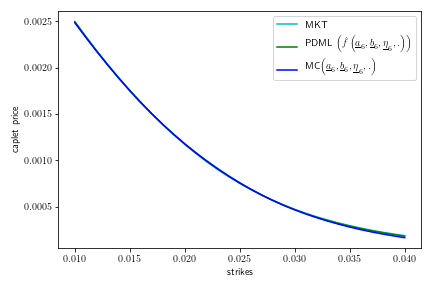}}
    \caption{More Robust Calibration Approach: Multiple Maturities. Prices From Best Seed PDML Network and MC at Optimized Parameter Sets Compared to Market Prices. (a) 1yr Maturity (b) 2yr Maturity (c) 3yr Maturity (d) 4yr Maturity (e) 5yr Maturity (f) 6yr Maturity. }
    \label{price_graphs_bestseed}
\end{figure}

In Figure \ref{price_graphs_bestseed}, 
we compare the price graphs of best seed PDML network and MC at optimized parameter sets
$\left(\underline{a}_{i},\underline{b}_{i},\underline{\eta}_{i}\right)$ 
against market prices across a range of strikes for $i=1, \dots, 6$. 
We can see that the optimized parameter sets from the best seed calibration approach reproduced 
market prices very well across range of strikes for all maturities considered.  

\begin{figure}
    \centering
    \subfigure[]{\includegraphics[width=0.49\textwidth]{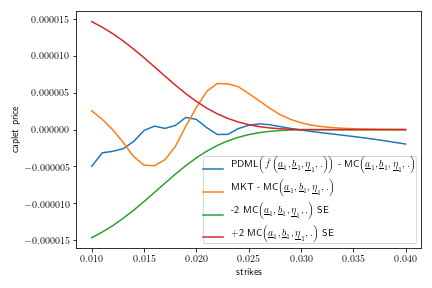}} 
    \subfigure[]{\includegraphics[width=0.49\textwidth]{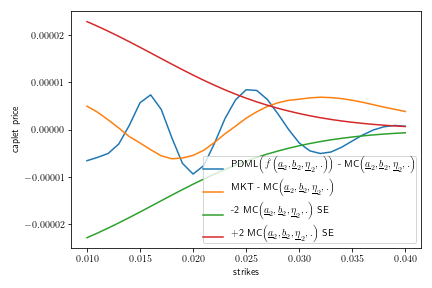}} 
    \subfigure[]{\includegraphics[width=0.49\textwidth]{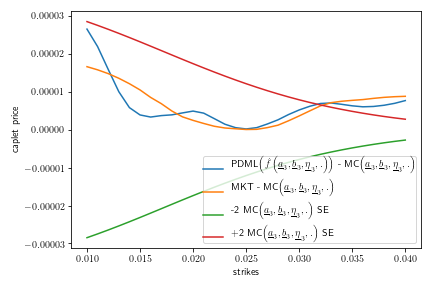}}
    \subfigure[]{\includegraphics[width=0.49\textwidth]{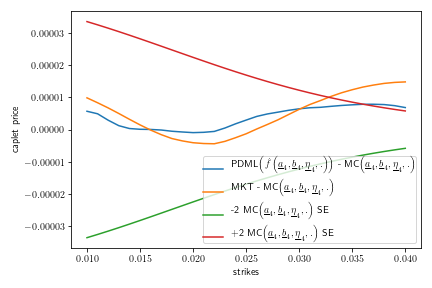}}
    \subfigure[]{\includegraphics[width=0.49\textwidth]{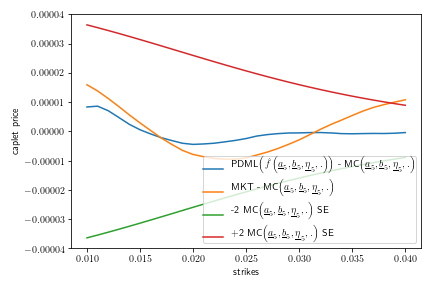}}
    \subfigure[]{\includegraphics[width=0.49\textwidth]{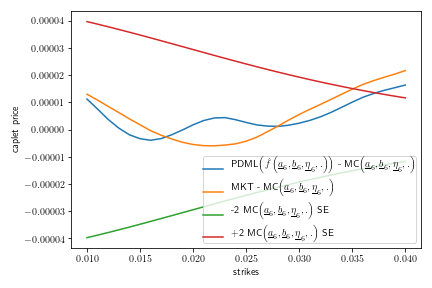}}
    \caption{More Robust Calibration Approach: Multiple Maturities. Comparing Price Differences: Best Seed PDML Network Prices vs MC Prices and vs Market Prices for Optimized Parameter Sets.
     (a) 1yr Maturity (b) 2yr Maturity (c) 3yr Maturity (d) 4yr Maturity (e) 5yr Maturity (f) 6yr Maturity. }
    \label{pricediff_graphs_bestseed}
\end{figure}

In Figure \ref{pricediff_graphs_bestseed}, 
we analyze prices more closely by looking at differences with respect to approximate ground truth, here represented by MC with enough samples.
PDML$\left(\hat{f}\left(\underline{a}_{i},\underline{b}_{i},\underline{\eta}_{i},.\right)\right)$ - MC$\left(\underline{a}_{i},\underline{b}_{i},\underline{\eta}_{i}, .\right)$ 
shows how well the PDML network replicated MC at the optimized parameter set $\left(\underline{a}_{i},\underline{b}_{i},\underline{\eta}_{i} \right)$.
MKT - MC$\left(\underline{a}_{i},\underline{b}_{i},\underline{\eta}_{i}, . \right)$ shows how well the optimized parameter set approximates 
calibration target prices (MKT) by considering MC with enough samples as approximate ground truth. 
$\pm 2$MC$\left(\underline{a}_{i},\underline{b}_{i},\underline{\eta}_{i}, . \right)$ SE standard error bars are provided 
as base line to assist in comparing differences. 
Except for 1 year maturity, the differences are well within 2 standard error bars for most of the strikes,
indicating that optimized parameter sets reproduced market prices very well and that the best seed PDML network agrees well with MC approximation of
ground truth at those parameter sets.

For the 1 year maturity, the PDML network is close to MC with enough samples and thus ground truth. 
Both are somewhat away from the target prices within the middle and right of the strike range. 
To be able to reprice those target prices in that region very well, 
one cannot assume that the model parameters are constant across the first year; 
one would need to calibrate to shorter maturities also, say 3M, 6M, and 9M, and 
calibrate model parameters for each quarter, 
or one would need to introduce a different and richer parametrization for the volatility term rather than the linear one used here. 
Thus, this corresponds to a model parametrization limitation rather than a 
limitation of the calibration and would be seen regardless of calibration method.

In Table \ref{table:3}, we present calibration results from the ensemble calibration approach. 
The blue color highlights the least max error and the red color highlights the highest max error 
among the five different seeds. 
We can see that ensemble results are close to least max error (blue) and sometimes better, for instance for 2 year maturity (highlighted in bold). 
\begin{table}[h!]
\centering
\begin{tabular}{ |c|c|c|c|c|c|c|c| }
\hline
Maturity & Metric & Seed:2476 & Seed:4548 & Seed:5670 & Seed:5818 & Seed:8642 & Ensemble \\
\hline
\multirow{3}{4em}{1 yr} & PDML fit error & 9.13E-11 & 6.49E-12  & 8.71E-12 & 1.11E-11  & 2.40E-12 & 4.28E-12  \\
& Model error & 8.90E-11  & 9.53E-12  & 1.42E-11   & 1.92E-11   & 9.62E-12 &  1.07E-11 \\
&max error & \textcolor{red}{9.13E-11}  & \textcolor{blue}{9.53E-12}  & 1.42E-11  & 1.92E-11  & 9.62E-12 & 1.07E-11   \\
\hline
\multirow{3}{4em}{2 yr} & PDML fit error & 2.56E-10  & 6.09E-11  & 1.70E-10  & 9.40E-12   & 5.28E-11 &  1.44E-11 \\
& Model error & 2.35E-10  & 5.92E-11  & 1.80E-10   & 3.35E-11   & 5.77E-11 &  3.16E-11 \\
&max error & \textcolor{red}{2.56E-10} & 6.09E-11  & 1.80E-10   & \textcolor{blue}{3.35E-11}  &  5.77E-11 & \textbf{3.16E-11}   \\
\hline
\multirow{3}{4em}{3 yr} & PDML fit error & 2.94E-10    &  2.78E-11 & 9.89E-12  & 3.08E-11  &  6.48E-11  & 2.22E-11  \\
& Model error & 3.05E-10  & 6.07E-11  & 4.20E-11  & 6.48E-11   & 1.22E-10 &  5.17E-11 \\
&max error &  \textcolor{red}{3.05E-10} & 6.07E-11  & \textcolor{blue}{4.20E-11}  &  6.48E-11  &  1.22E-10  &  5.17E-11 \\
\hline
\multirow{3}{4em}{4 yr} & PDML fit error & 1.46E-09   &  4.97E-11  & 1.55E-10 &  2.00E-11   & 7.78E-12 & 1.50E-11   \\
& Model error &  1.52E-09  &  9.27E-11  &  2.02E-10  &  7.40E-11   & 6.09E-11  &  6.47E-11 \\
&max error &  \textcolor{red}{1.52E-09}  &   9.27E-11 &  2.02E-10  &  7.40E-11   &  \textcolor{blue}{6.09E-11}   & 6.47E-11  \\
\hline
\multirow{3}{4em}{5 yr} & PDML fit error & 3.30E-10  & 1.46E-10  & 2.50E-10  & 5.48E-11  & 3.86E-11  & 3.28E-11 \\
& Model error & 3.51E-10  & 1.94E-10  & 2.98E-10   &  8.17E-11 &  8.00E-11  &  8.45E-11 \\
&max error &  \textcolor{red}{3.51E-10} &  1.94E-10  &  2.98E-10  &  8.17E-11 &  \textcolor{blue}{8.00E-11} &   8.45E-11 \\
\hline
\multirow{3}{4em}{6 yr} & PDML fit error & 7.33E-10  & 1.17E-11 & 9.20E-11   & 3.07E-10   &  9.35E-10  & 3.88E-12  \\
& Model error & 7.73E-10  & 8.99E-11  & 1.77E-10  & 3.78E-10   & 1.02E-09  & 9.78E-11 \\
&max error & 7.73E-10  & \textcolor{blue}{8.99E-11}  &  1.77E-10 &  3.78E-10  &  \textcolor{red}{1.02E-09} &  9.78E-11 \\
\hline
\end{tabular}
\caption{More Robust Calibration Approach: Multiple Maturities. Calibration Results With Ensemble Approach for Multiple Maturities.}
\label{table:3}
\end{table}

\begin{figure}
    \centering
    \subfigure[]{\includegraphics[width=0.49\textwidth]{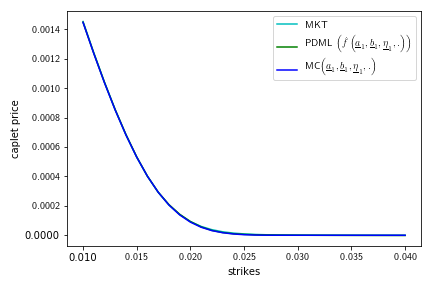}} 
    \subfigure[]{\includegraphics[width=0.49\textwidth]{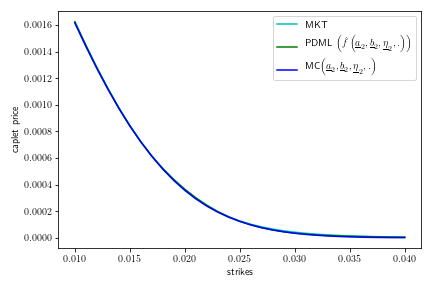}} 
    \subfigure[]{\includegraphics[width=0.49\textwidth]{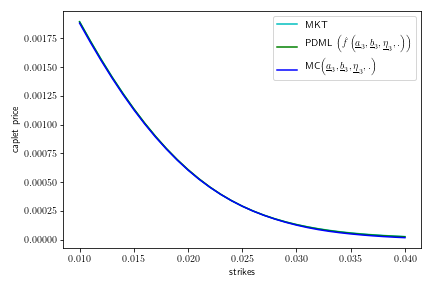}}
    \subfigure[]{\includegraphics[width=0.49\textwidth]{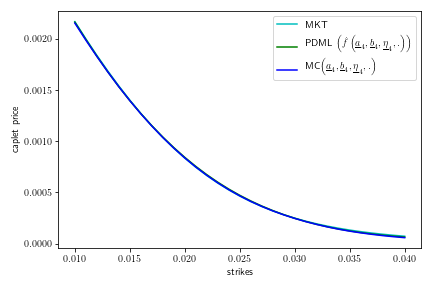}}
    \subfigure[]{\includegraphics[width=0.49\textwidth]{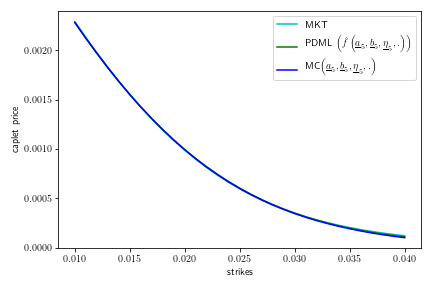}}
    \subfigure[]{\includegraphics[width=0.49\textwidth]{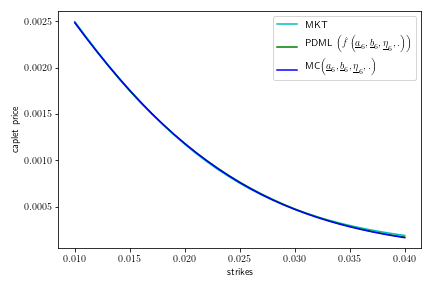}}
    \caption{More Robust Calibration Approach: Multiple Maturities. Comparing Ensemble PDML Network Prices vs MC Prices and vs Market Prices for Optimized Parameter Sets.
     (a) 1yr Maturity (b) 2yr Maturity (c) 3yr Maturity (d) 4yr Maturity (e) 5yr Maturity (f) 6yr Maturity.}
    \label{price_graphs_ensemble}
\end{figure}

\begin{figure}
    \centering
    \subfigure[]{\includegraphics[width=0.49\textwidth]{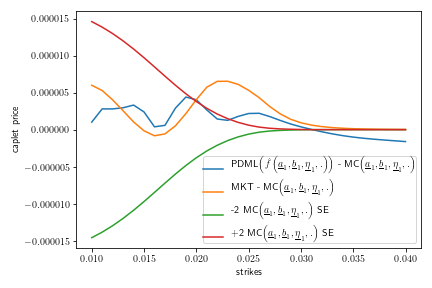}} 
    \subfigure[]{\includegraphics[width=0.49\textwidth]{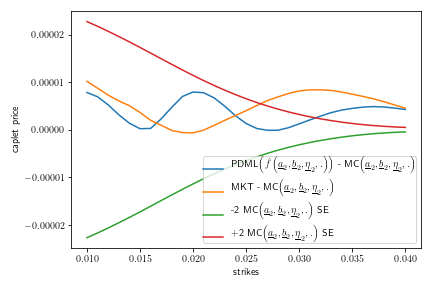}} 
    \subfigure[]{\includegraphics[width=0.49\textwidth]{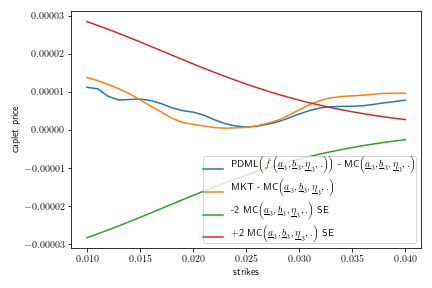}}
    \subfigure[]{\includegraphics[width=0.49\textwidth]{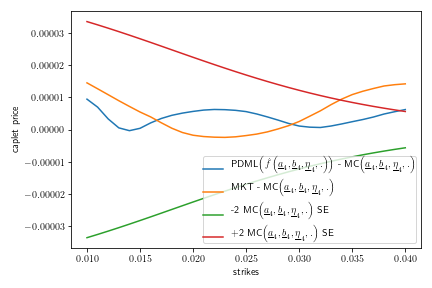}}
    \subfigure[]{\includegraphics[width=0.49\textwidth]{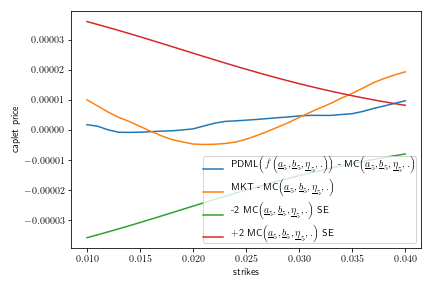}}
    \subfigure[]{\includegraphics[width=0.49\textwidth]{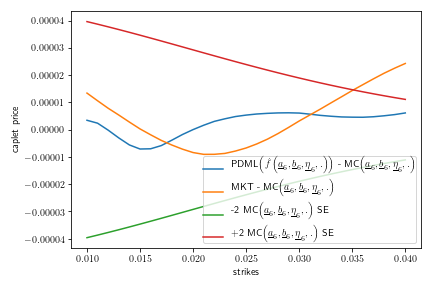}}
    \caption{More Robust Calibration Approach: Multiple Maturities. Comparing Price Differences: Ensemble PDML Network Prices vs MC Prices and vs Market Prices for Optimized Parameter Sets.
     (a) 1yr Maturity (b) 2yr Maturity (c) 3yr Maturity (d) 4yr Maturity (e) 5yr Maturity (f) 6yr Maturity. }
    \label{pricediff_graphs_ensemble}
\end{figure}

In Figures \ref{price_graphs_ensemble} and \ref{pricediff_graphs_ensemble}, 
we can see that ensemble approach has similar results as best seed approach in 
Figures \ref{price_graphs_bestseed} and \ref{pricediff_graphs_bestseed}. 
In Table \ref{table:4}, we can see that the accuracy of best seed and ensemble approaches is similar. 
As both approaches have similar accuracy, we chose the best seed approach for our calibration application 
as best seed approach has smaller computational cost  than the ensemble approach. 
Nonetheless, we believe ensemble approach could be a good alternative approach
 for calibration with PDML technique, in particular also for other models and set-ups.  

\begin{table}[h!]
\centering
\begin{tabular}{|c|c|c|}
\hline
Maturity \textbackslash Approach & Best seed & Ensemble \\
\hline
1yr & 9.53E-12 & 1.07E-11 \\
\hline
2yr & 2.47E-11 & 3.16E-11 \\
\hline
3yr & 2.44E-11 & 5.17E-11 \\
\hline
4yr & 3.79E-11 & 6.47E-11 \\
\hline
5yr & 4.09E-11 & 8.45E-11 \\
\hline
6yr & 1.09E-10 & 9.78E-11 \\
\hline
\end{tabular}
\caption{More Robust Calibration Approach: Multiple Maturities. Comparison of Max Error of Best Seed and Ensemble Approaches.}
\label{table:4}
\end{table}

\section{Conclusion} \label{conclusion}

In this paper, we demonstrated how Parametric Differential Machine Learning (PDML) can produce accurate enough parametric pricing surrogates 
for conditional expectations of functionals 
from samples of the functional and sample-wise derivatives of samples with respect to parameters
for appopriately sampled parameters  
as generated by differentiable simulation and computation 
from text input close to mathematical notation.
If different model parameters lead to prices of different magnitudes, adaptive parameter sampling 
is used to improve relative approximation for parameter ranges 
for which the magnitude of the price is smaller, leading to better approximation 
and calibration in such parameter ranges. 
These parametric pricing surrogates can stand in for specialized solvers for purposes such as calibration and speed up the calibration. 
Since the sampling, the construction of surrogates, and the optimization for calibration is impacted by randomness; one can generate
different surrogates and different optimized parameter sets from different seeds. Once a ground truth indicator has been implemented, 
one can then pick surrogates that approximate ground truth well and approximate calibration target prices, at least close to the 
optimized parameter sets. 
Such surrogates and optimization methods will be more accurate and more reliable than what would be obtained with a single seed. 

We demonstrated parametric pricing and calibration for interest rate caplets for Cheyette Models
with benchmark rate volatility specification with an additional stochastic volatility factor and obtain 
good results. 

As such, the demonstrated parametric differential machine learning approach to parameteric
pricing and calibration gives promising results without the need to implement simplified 
and/or specialized solvers for the calibration instruments or the need to implement specialized 
simulation methodologies. We believe that PDML is thus a good choice for parametric 
pricing and calibration. Due to its speed, it will allow tests and computations that have been
so far impossible with other techniques. In particular, with PDML one can now find several 
parameter sets that calibrate equally well to given calibration targets and consider the model risk impact
of picking just one calibrated parameter set for pricing and risk management 
versus comparing robustness of pricing and risk management over all these 
alternative optimized parameter sets.

{\bf Acknowledgments:}

The authors thank Vijayan Nair for his comments and suggestions regarding this research. 
Bernhard Hientzsch thanks Todd Story for prior discussions
and collaboration on calibration of Cheyette Models with short
rate volatility specifications with Monte Carlo pricers.   
Any opinions, findings and conclusions or recommendations expressed in this material are those of the
authors and do not necessarily reflect the views of Wells Fargo Bank, N.A., its parent
company, affiliates and subsidiaries.

\newpage

\bibliographystyle{alpha}
\bibliography{../Bibliographies/dml}

\newcommand{\etalchar}[1]{$^{#1}$}
\begin{thebibliography}{CGP{\etalchar{+}}20}

\bibitem[And01]{andreasen2001turbo}
Jesper Andreasen.
\newblock Turbo charging the {C}heyette model.
\newblock {\em Available at SSRN 1719142}, 2001.

\bibitem[AP10]{andersenpiterbarginterest}
L~Andersen and V~Piterbarg.
\newblock {\em Interest Rate Modeling}.
\newblock Atlantic Financial Press: London, 2010.

\bibitem[CGP{\etalchar{+}}19]{cyr2019robustarxiv}
Eric~C. Cyr, Mamikon~A. Gulian, Ravi~G. Patel, Mauro Perego, and Nathaniel~A.
  Trask.
\newblock Robust training and initialization of deep neural networks: An
  adaptive basis viewpoint.
\newblock {\em arXiv preprint arXiv:1912.04862}, 2019.

\bibitem[CGP{\etalchar{+}}20]{cyr2020robust}
Eric~C Cyr, Mamikon~A Gulian, Ravi~G Patel, Mauro Perego, and Nathaniel~A
  Trask.
\newblock Robust training and initialization of deep neural networks: An
  adaptive basis viewpoint.
\newblock In {\em Mathematical and Scientific Machine Learning}, pages
  512--536. PMLR, 2020.

\bibitem[Hen10]{MH_multicurve}
Marc~P.A. Henrard.
\newblock The irony in the derivatives discounting, {P}art {II}: {T}he crisis.
\newblock {\em Wilmott Journal}, 2:301--316, 2010.
\newblock Also available at SSRN.

\bibitem[HS20a]{huge2020differential}
Brian Huge and Antoine Savine.
\newblock Differential machine learning.
\newblock {\em arXiv preprint arXiv:2005.02347}, 2020.

\bibitem[HS20b]{huge2020differentialrisk}
Brian Huge and Antoine Savine.
\newblock Differential machine learning: {T}he shape of things to come.
\newblock {\em Risk Magazine}, 2020.

\bibitem[JJ04]{joshi2004c++}
Mark~S Joshi and Mark~Suresh Joshi.
\newblock {\em C++ design patterns and derivatives pricing}, volume~1.
\newblock Cambridge University Press, 2004.

\bibitem[JWCJ13]{jia2013improved}
Guanbo Jia, Yong Wang, Zixing Cai, and Yaochu Jin.
\newblock An improved ($\mu$+ $\lambda$)-constrained differential evolution for
  constrained optimization.
\newblock {\em Information Sciences}, 222:302--322, 2013.

\bibitem[KB14]{kingma2014adam}
Diederik~P Kingma and Jimmy Ba.
\newblock Adam: A method for stochastic optimization.
\newblock {\em arXiv preprint arXiv:1412.6980}, 2014.

\bibitem[Nau11]{naumann2011art}
Uwe Naumann.
\newblock {\em The art of differentiating computer programs: {A}n introduction
  to algorithmic differentiation}.
\newblock SIAM, 2011.

\bibitem[{\"O}ge20]{ogetbil2020extensions}
Orcan {\"O}getbil.
\newblock Extensions of dupire formula: Stochastic interest rates and
  stochastic local volatility.
\newblock {\em arXiv preprint arXiv:2005.05530}, 2020.

\bibitem[{\"O}GH20]{ogetbil2020calibrating}
Orcan {\"O}getbil, Narayan Ganesan, and Bernhard Hientzsch.
\newblock Calibrating local volatility models with stochastic drift and
  diffusion.
\newblock {\em arXiv preprint arXiv:2009.14764}, 2020.

\bibitem[{\"O}GH22]{ogetbil2022calibrating}
Orcan {\"O}getbil, Narayan Ganesan, and Bernhard Hientzsch.
\newblock Calibrating local volatility models with stochastic drift and
  diffusion.
\newblock {\em International Journal of Theoretical and Applied Finance},
  25(02):2250011, 2022.

\bibitem[{\"O}H22]{ogetbil2022flexible}
Orcan {\"O}getbil and Bernhard Hientzsch.
\newblock A flexible commodity skew model with maturity effects.
\newblock {\em arXiv preprint arXiv:2212.07972}, 2022.

\bibitem[PSL05]{price2005differential}
Kenneth Price, Rainer~M Storn, and Jouni~A Lampinen.
\newblock {\em Differential Evolution: A Practical Approach to Global
  Optimization (Natural Computing Series)}.
\newblock Springer-Verlag, 2005.

\bibitem[SA21]{savine2021modern}
Antoine Savine and Jesper Andreasen.
\newblock {\em Modern Computational Finance: {S}cripting for Derivatives and
  xVA}.
\newblock John Wiley \& Sons, 2021.

\bibitem[Sav18]{savine2018modern}
Antoine Savine.
\newblock {\em Modern computational finance: AAD and parallel simulations}.
\newblock John Wiley \& Sons, 2018.

\bibitem[Sch16]{schlenkrich2016quasigaussian}
Sebastian Schlenkrich.
\newblock Quasi-{G}aussian model in {Q}uantlib.
\newblock {\em https://www.quantlib.org/slides/qlum16d/schlenkrich.pdf}, 2016.

\end{thebibliography}

\end{document}